\documentclass[11pt]{article}
\usepackage{moriond,epsfig}
\usepackage{wrapfig}

\def\ptt  {$p^2_t$}
\def\lambdazero    {$\Lambda$}
\def\antilambda    {$\overline{\Lambda}$}

\def\be{\begin{equation}}
\def\ee{\end{equation}}
\def\bea{\begin{eqnarray}}
\def\eea{\end{eqnarray}}

\newcommand{\ksp}{K^{0}_{S}\> p}
\newcommand{\kspb}{K^{0}_{S}\> \bar{p}}
\newcommand{\ksppb}{K^{0}_{S}\> p\>(\bar{p})}
\newcommand{\coll}{Collaboration}
\newcommand{\etal}{ {\it et al.,} }

\newcommand{\BR}{{\cal B}}
\newcommand{\kskl}{K_S^0 K_L^0}
\newcommand{\psp}{\psi^\prime}
\newcommand{\jpsi}{J/\psi}

\newcommand{\ppb}{p\overline{p}}
\newcommand{\aab}{\Lambda\overline{\Lambda}}
\newcommand{\ksks}{K^0_S K^0_S}
\newcommand{\jpsito}{J/\psi \rightarrow }
\newcommand{\pspto}{\psp \rightarrow }
\newcommand{\chicJto}{\chi_{cJ} \rightarrow }

\def\babar{\mbox{\slshape B\kern-0.1em{\smaller A}\kern-0.1em B\kern-0.1em{\smaller A\kern-0.2em R}}}
\newcommand{\etapr}{\ensuremath{\eta^\prime}}
\def\Kstarz  {\ensuremath{K^{*0}}}
\def\Kstar   {\ensuremath{K^*}}
\newcommand{\etaKst}{\ensuremath{B\to\eta K^{*}}}
\newcommand{\etapKst}{\ensuremath{B\to\etapr K^{*}}}
\newcommand{\etapKz}{\ensuremath{B^0\to\etapr K^0}}
\def\piz   {\ensuremath{\pi^0}}
\def\KS{\ensuremath{K^0_{\scriptscriptstyle S}}} 

\def\BABAR{{\sc $B$a$B$ar~}}
\def\babar{{\sc $B$a$B$ar~}}
\def\sss{\scriptscriptstyle}
\def\barpd{{\raise.35ex\hbox{${\sss (}$}}--{\raise.35ex\hbox{${\sss )}$}}}
\def\dbarp{\hbox{$D^{0}$\kern-1.3em\raise1.5ex\hbox{\barpd}}}

\def \gluino {\tilde{g}}

\def \squark {\tilde{q}}

\def\scr{\scriptstyle}

\def\jpsi{{J/\psi}}
\def\ifb{fb^{-1}}

\def\gsim{\mathrel{\rlap{\raise.4ex\hbox{$>$}} {\lower.6ex\hbox{$\sim$}}}}
\def\lsim{\mathrel{\rlap{\raise.4ex\hbox{$<$}} {\lower.6ex\hbox{$\sim$}}}}

\include{def}

\begin{document}
\vspace*{4cm}
\pagenumbering{arabic}
\pagestyle{plain}
\title{Rencontres du Moriond 2004 - QCD and Hadronic Interactions\\EXPERIMENTAL SUMMARY}

\author{Boaz~Klima}

\address{Fermilab \\
   Batavia, Illinois 60510, USA \\
(klima@fnal.gov)
}
\maketitle

\begin{abstract}
Highlights of the experimental results presented at the $39^{th}$ Rencontres du Moriond on ``QCD and High Energy Hadronic Interactions'', which was held in La Thuile, Italy on Mar.~28 - Apr.~4, 2004 are briefly summarised.
\end{abstract}

\section{Introduction}

In this paper I attempt to briefly present the main experimental results that were shown at the $39^{th}$ Rencontres du Moriond on ``QCD and High Energy Hadronic Interactions''. These new and impressive results on a variety of subjects were split into six general categories: Multi-Quark States, Heavy Ion Interactions, Core QCD, Heavy Flavors, Top and Higgs, and New Phenomena Searches. Since I had to summarize 65 experimental talks in a limited space (and time) I could not mention all the results here; some interesting contributions were left out. More details on any of the results presented at this conference can be found in the contributions made to the Proceeedings by the individual speakers.

\section{Multi-Quark States}
\label{sec:mqs}

Recently, interest in light hadron spectroscopy has been considerably revived
by the observation of baryonic resonances \cite{fixed,ks,NA49} 
compatible with their interpretation in terms of pentaquark states, 
i.e. bound states of four quarks and an antiquark. While almost all hadronic states observed previously can be interpreted in terms of baryons or mesons, the theory of Quantum Chromo Dynamics (QCD) does not preclude the existence of other 
color neutral quark combinations such as tetraquarks (qq\=q\=q), pentaquarks
(qqqq\=q), etc., or states including gluons such as glueballs (gg) or
hybrids (q\=qg). Potential glueball candidates have been discussed for many years without firm conclusions.

\begin{wrapfigure}{r}{6cm}
\vspace*{-0.3in}
  \centerline{\psfig{figure=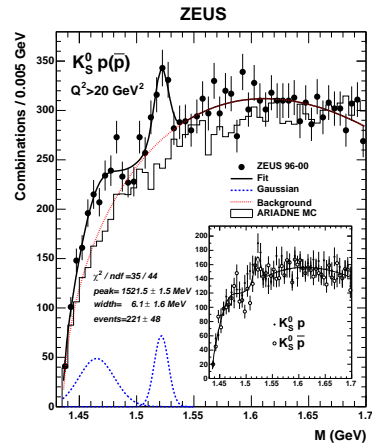,height=6.0cm}}
  \caption{Invariant-mass spectrum for the $\ksppb$ channel. The result of a fit (solid), the Gaussian components (dashed) and the background (dotted) are shown.
    \label{fig:theta_dist}}
\end{wrapfigure}

Recent results on light hadron spectroscopy were reported~\cite{geiser},
with special emphasis on the evidence for a narrow baryonic state decaying
to $\ksp$, compatible with the pentaquark state $\theta^+$ 
observed by fixed target experiments.
The data were collected with the ZEUS detector at HERA using an integrated 
luminosity of 121 pb$^{-1}$. The analyses were performed in the
central rapidity region of inclusive deep inelastic scattering at
an $ep$ center-of-mass energy of 300--318 GeV. Evidence for a narrow resonance in the $\ksp$ invariant mass 
spectrum is obtained, with mass $1521.5 \pm 1.5(stat)^{+2.8}_{-1.7}(syst)$
and width consistent with the experimental resolution (see Fig.~\ref{fig:theta_dist}). If the $\ksp$ part of the signal is identified with the strange pentaquark $\theta^+$, the $\kspb$ part (shown in inset) is the first evidence for its antiparticle, $\bar \theta^-$.

\begin{figure}[hbt]
\begin{center}
\includegraphics[width=0.3\textwidth]{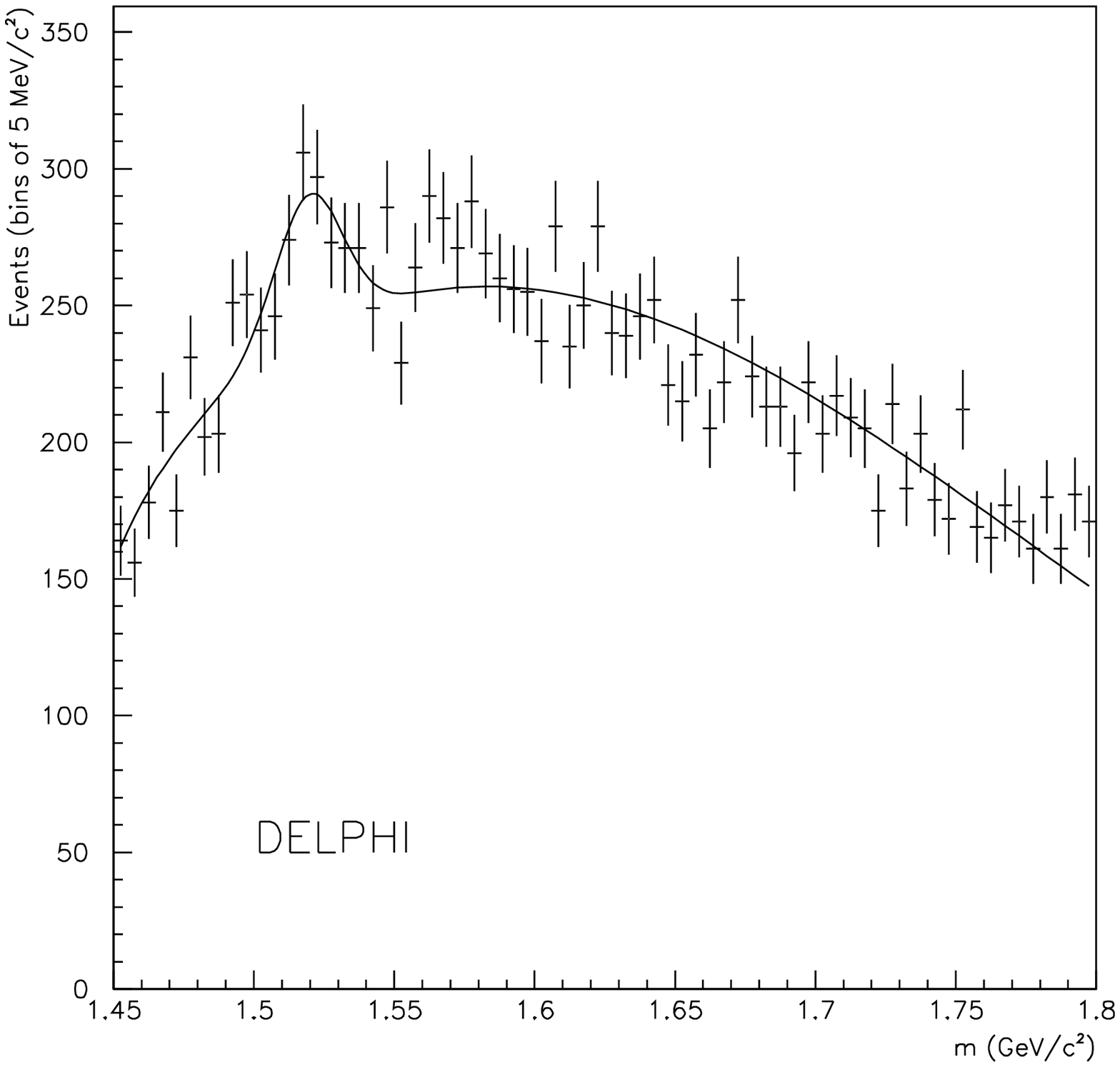}
\includegraphics[width=0.3\textwidth]{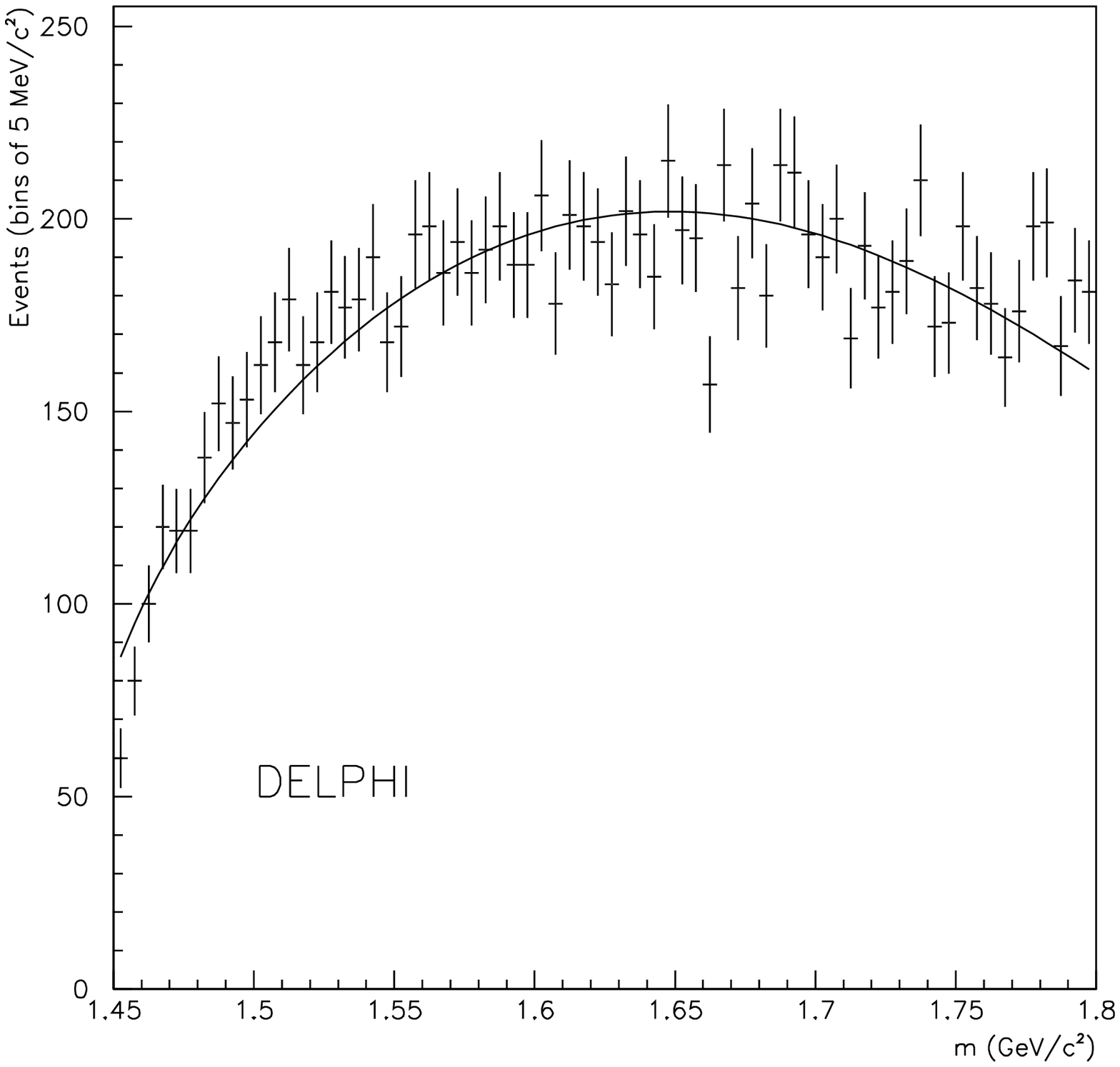}
\caption{The $pK^-$ (left) and the $pK^0$ (right) invariant mass spectrum as observed by DELPHI.}
\label{fig:penta}
\end{center}
\end{figure}

DELPHI has performed a similar search~\cite{wengler} for a narrow baryonic resonance in the proton--kaon system. If such a state is
produced at LEP1 like an ordinary baryon, its production rate per
hadronic event should be similar to that of the
$\Lambda\left(1520\right)$, which is $0.0224 \pm 0.0027$. The left
plot of Figure~\ref{fig:penta} shows the $pK^-$ invariant mass
observed by DELPHI in the data taken at the $Z$-peak. A resonance
consistent with the $\Lambda\left(1520\right)$ is observed,
demonstrating the ability of the experiment to reconstruct such
states with the available data sets. No resonance structure is
observed in the $pK^0$ invariant mass spectrum shown on the right
hand side of Figure~\ref{fig:penta}.

The NA49 Collaboration has presented~\cite{tsusa} the results of resonance searches in $\Xi^{-} \pi^{-}$, $\Xi^{-} \pi^{+}$, $\overline{\Xi}^{+}\pi^{-}$ and $\overline{\Xi}^{+}\pi^{+}$ invariant mass spectra in proton-proton collisions at $\sqrt{s}=$17.2~GeV. A narrow state was observed in $\Xi^{-} \pi^{-}$ spectra with mass of  1.862 $\pm$ 0.002 GeV/$c^{2}$ and width below the detector resolution of about 0.018 GeV/$c^{2}$.
This state is identified as a candidate for the hypothesized exotic $\Xi_{5}^{--}$ baryon with S = $-2$, I = $\frac{3}{2}$ and a quark content of ($dsds\bar{u}$). $\Xi^{-}\pi^{+}$  and the corresponding antiparticle spectra show indication of enhancements at the same mass.

\begin{wrapfigure}{r}{6cm}
\vspace*{-0.7in}
  \centerline{\psfig{figure=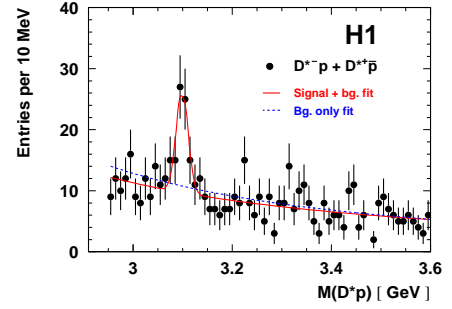,height=6.5cm}}
\vspace*{-0.3in}
\caption{$M (D^*p)$ distribution for $D^*p$ combinations in DIS compared with fits for signal plus background (solid) and background only (dashed).
\label{cpq_fit}}
\end{wrapfigure}

Evidence for an exotic anti-charmed baryon state, which has been observed by H1, was presented~\cite{lipka}. Their data show a narrow resonance in the $D^*p$ invariant mass combination at 3099$\pm$3$(stat.)$$\pm$5$(syst.)$ MeV (Fig. \ref{cpq_fit}). The resonance is interpreted as an anti-charmed baryon with minimal constituent quark content $uudd\bar{c}$ together with its charge conjugate. Such a signal is not observed in a similar preliminary ZEUS analysis.

\begin{figure}[hbt]
\begin{center}
\psfig{figure=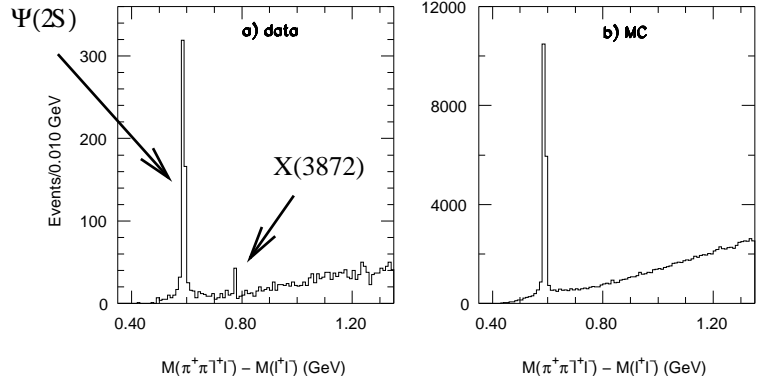,height=5cm}
\caption{Distribution of $m_{J/\psi\pi\pi}$ in Belle's data (a) and Monte Carlo.}
\label{3872}
\end{center}
\end{figure}

The Belle collaboration reported~\cite{mikami} on the first observation of X(3872) from $J/\psi\ \pi^+ \pi^-$ final state as shown in Figure \ref{3872}. It is interesting to note that this particle's mass has a value very close to the sum of $D^0$ and $D^{*0}$ masses. They also reported on first observation of the new $D_{sJ}$ from B decay, $D_{sJ}(2457)$ radiative decay and $D_{sJ}(2457)$ dipion decay. They measured masses of newly found $D_{sJ}$ and newly found broad $D^{**}$ states.

\begin{figure}[hbt]
\begin{center}
\includegraphics[width=0.3\textwidth]{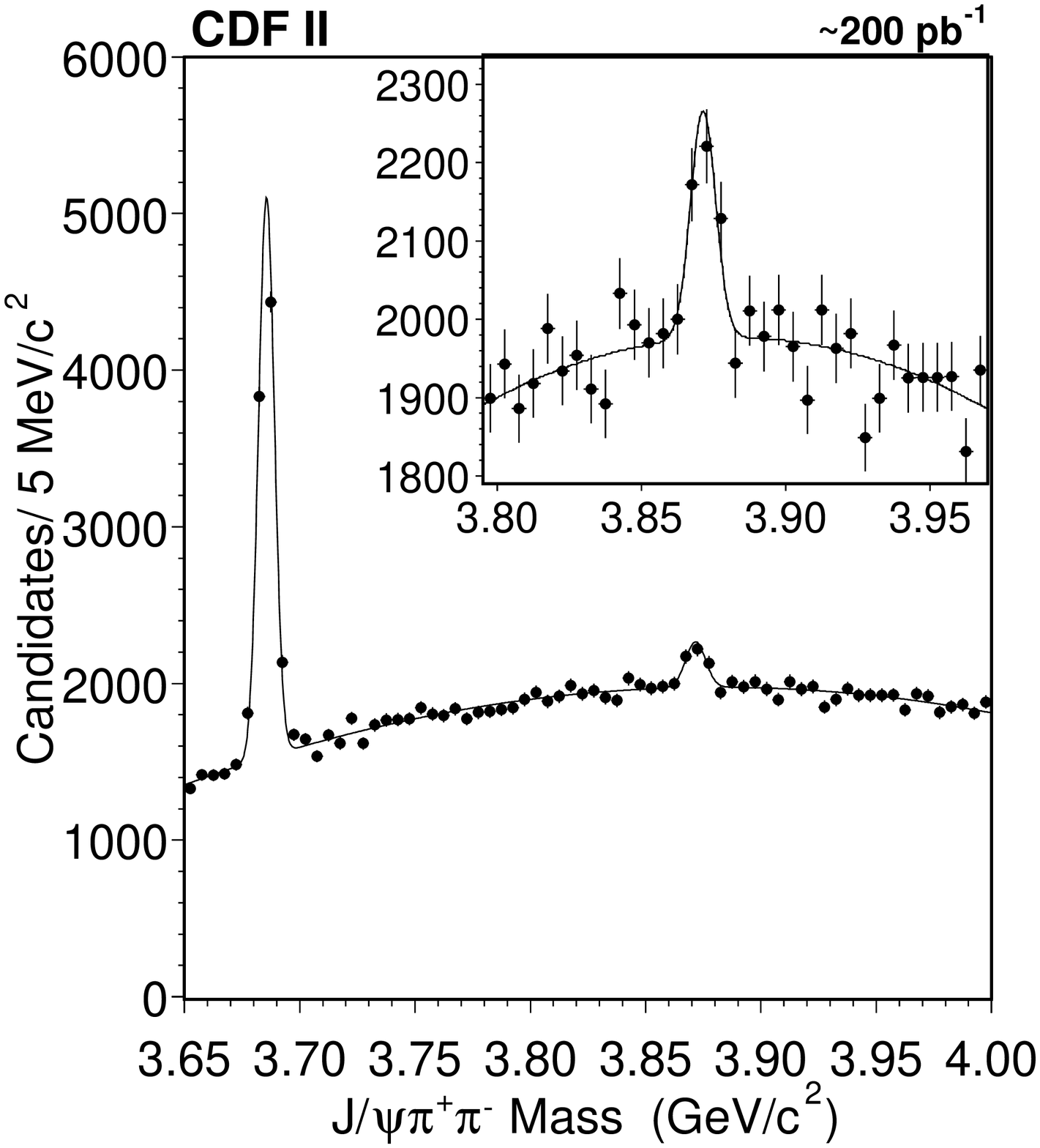}
\includegraphics[width=0.35\textwidth]{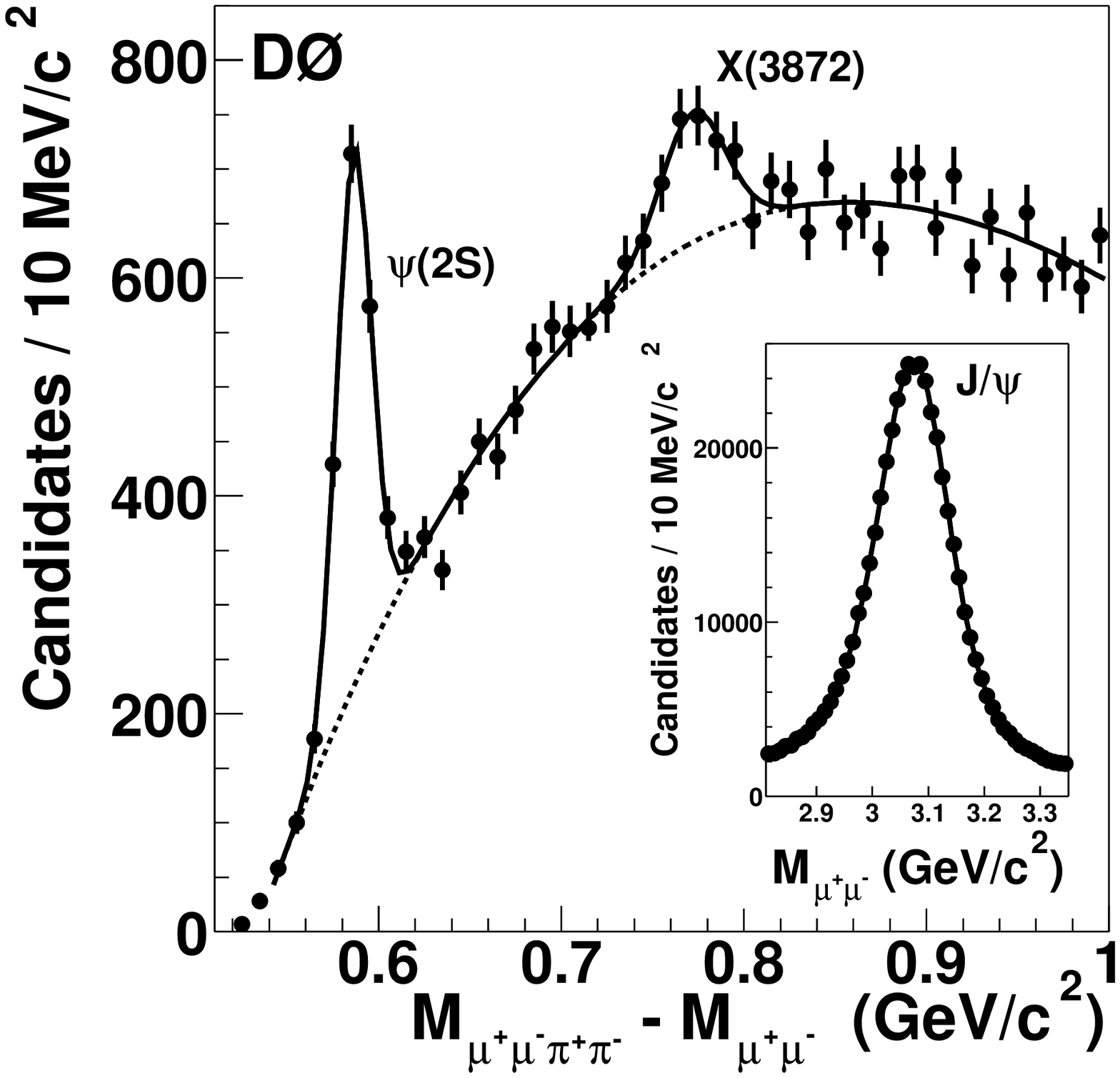}
\caption{$J/\Psi \pi^+\pi^-$ mass distribution for CDF (left) and D\O\ (right).
\label{fig:xmass}}
\end{center}
\end{figure}

The X(3872) signal from Belle has been confirmed by \BABAR \cite{calderini} and by both Tevatron detectors~\cite{campanelli} (Fig.~\ref{fig:xmass}). CDF has observed a 11$\sigma$ signal, with mass $M(X) = 3871.3\pm 0.7\pm 0.4$ MeV, while D\O\ has a 4.4$\sigma$ signal, with mass difference $M(X)-M(\Psi(2S)) = 766.4 \pm 3.5 \pm 3.9$ MeV, consistent with the other observations. The production characteristics of X in the D\O\ data seem to be similar to that of $\Psi$(2S).

\section{Heavy Ion Interactions}
\label{sec:hi}

Nucleus-nucleus collisions at relativistic energies aim at the 
study of the fundamental theory of the strong interaction, 
Quantum Chromo Dynamics (QCD), at extreme energy densities. 
The main goal of this physics program is the production and 
study under laboratory conditions of the plasma of quarks and 
gluons (QGP). The QGP is a deconfined and chirally symmetric 
state of strongly interacting matter predicted by QCD calculations 
on the lattice for values of the energy density 
five times larger than those found in the nuclear ground state.
The combination of high center-of-mass energies and large nuclear 
systems in the initial-state of heavy-ion reactions provides, furthermore, 
favorable conditions for the study of the (non-linear) parton dynamics 
at small values of (Bjorken) fractional momentum $x$. In this regime
(often  dubbed ``Color Glass Condensate'', CGC), higher-twist effects are expected to saturate the rapidly increasing 
density of ``wee'' gluons observed at small-$x$ in the hadronic wave 
functions which would, otherwise, violate the unitarity limit of the theory.

\begin{wrapfigure}{r}{6cm}
\vspace*{-0.3in}
  \centerline{\psfig{figure=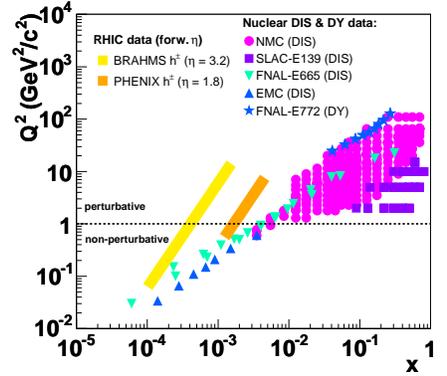,height=5.0cm}}
\vspace*{-0.1in}
\caption{Kinematical range in the $x$-$Q^2$ plane probed in nuclear DIS
and DY processes, and in $d+Au$ at forward rapidities at RHIC.
\label{fig:Rcp_brahms}}
\end{wrapfigure}

BRAHMS~\cite{denterria} results on high $p_T$ 
charged hadron production at pseudorapidities $\eta$ = 3.2 and 
$\eta$ = 1.8 (corresponding to $x\approx$ $\mathcal{O}$(10$^{-4}$) 
and $\mathcal{O}$(10$^{-3}$) respectively) show a suppression instead of 
an enhancement as found at $\eta$ = 0. 
This is the first time that the nuclear PDFs are probed 
at such small values of $x$ in the {\it perturbative} domain 
($Q^2\approx p_T^2>$ 1 GeV$^2/c^2$) (Fig.~\ref{fig:Rcp_brahms}). BRAHMS result seems to indicate that a less significant 
amount of gluon ``shadowing'' in this range. Whether this larger suppression is due to soft physics (the global $dN/dy$ distributions in 
lower energy $p+A$ collisions are also found to be depleted at 
forward $\eta$) or a genuine CGC effect, is still matter of 
discussion at this point.

\begin{figure}[htbp]
\begin{center}
\epsfig{file=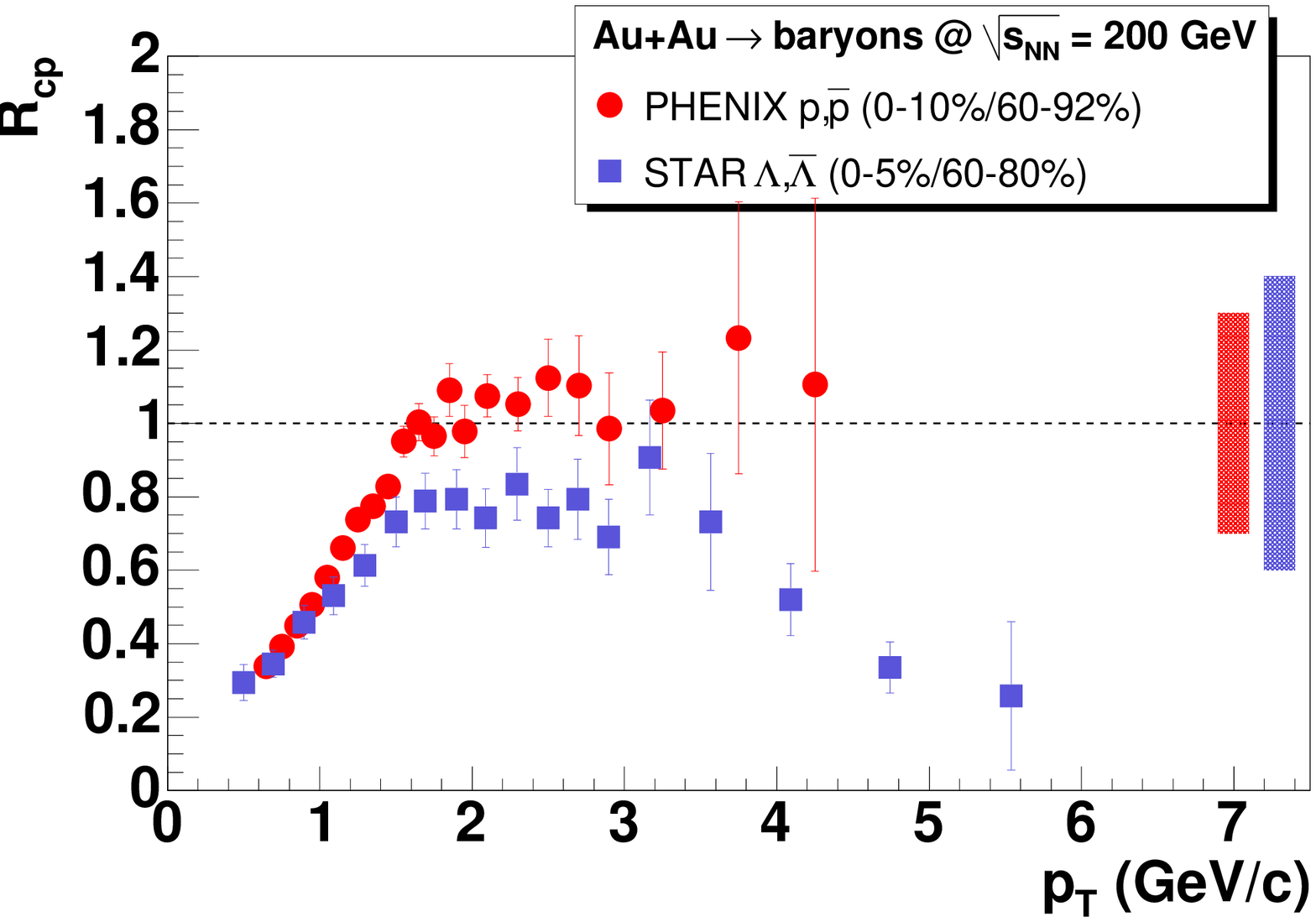,height=5.2cm}
\epsfig{file=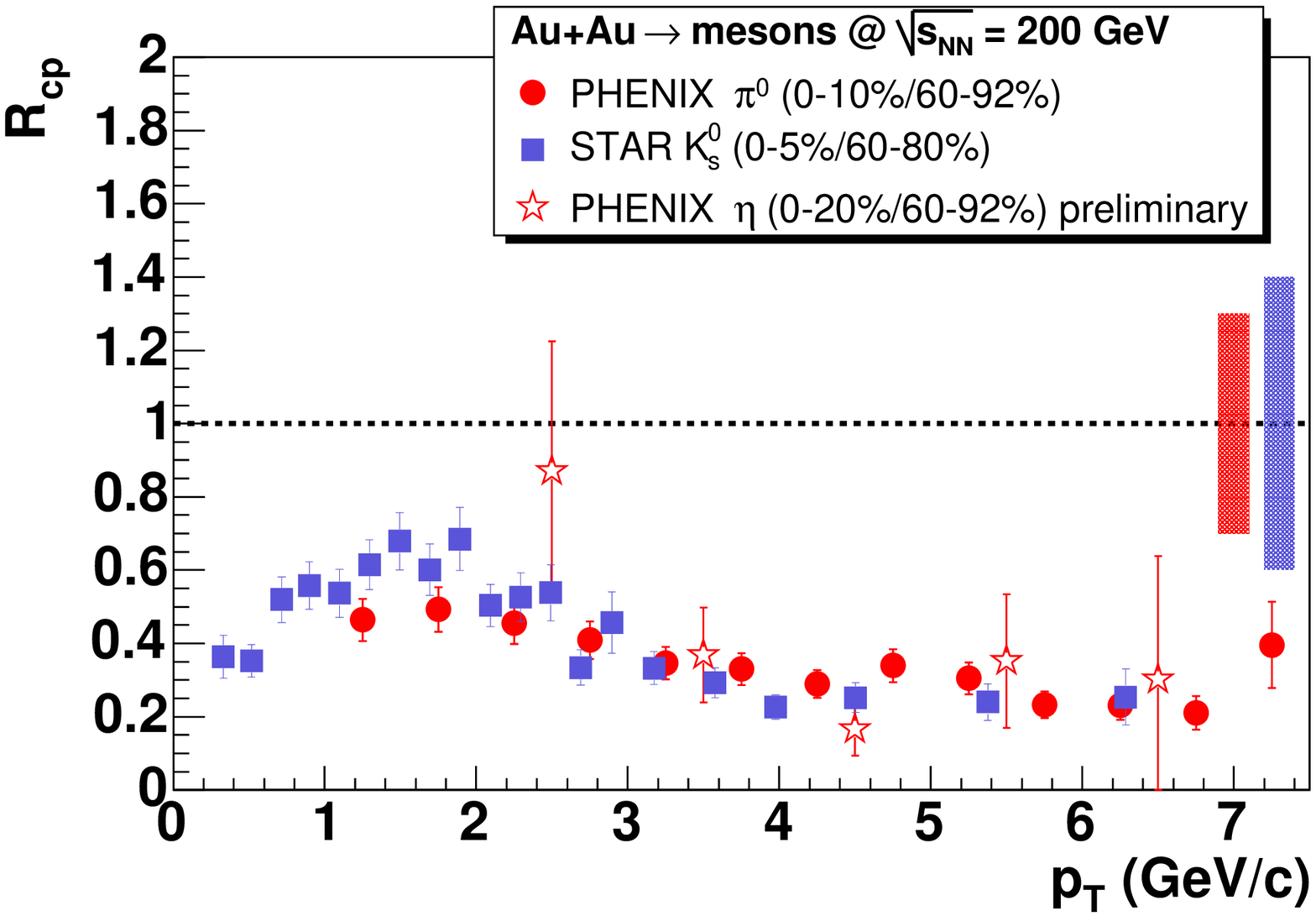,height=5.2cm}
\end{center}
\vspace{-0.5cm}
\caption[]{Ratio of central over peripheral $N_{coll}$ scaled yields, $R_{cp}$, as a function of $p_{T}$ for baryons (left): $(p+\bar{p})/2$ (dots) and 
$(\Lambda+\bar{\Lambda})/2$ (squares); and for mesons (right): $\pi^0$ (dots), 
$K^0_s$ (squares) and $\eta$ (stars); measured in $Au-Au$ collisions at 200 GeV. The shaded bands indicate the associated fractional normalization uncertainties.}
\label{fig:flavor_dep}
\end{figure}

An intriguing result of the RHIC program~\cite{denterria} is the different suppression 
pattern of baryons and mesons at moderately high $p_{T}$.
Figure~\ref{fig:flavor_dep} shows the $N_{coll}$ scaled central 
to peripheral yield ratios, $R_{cp}$, for baryons (left) and mesons (right). In the range 
$p_T\approx$ 2 -- 4 GeV/$c$ the (anti)protons are not suppressed, at variance with the pions which are reduced 
by a factor of 2 -- 3. The resulting baryon/meson$\sim$0.8 ratio is clearly at odds with the ``perturbative'' $\sim$0.2 
ratio measured in $p+p$ or $e^{+}e^{-}$ collisions. Since such a 
particle composition is inconsistent with standard fragmentation 
functions, there must be an additional non-perturbative mechanism for 
baryon production in central Au-Au reactions in this intermediate $p_T$ range. 

\begin{figure}[hbt]
\begin{center}
\psfig{figure=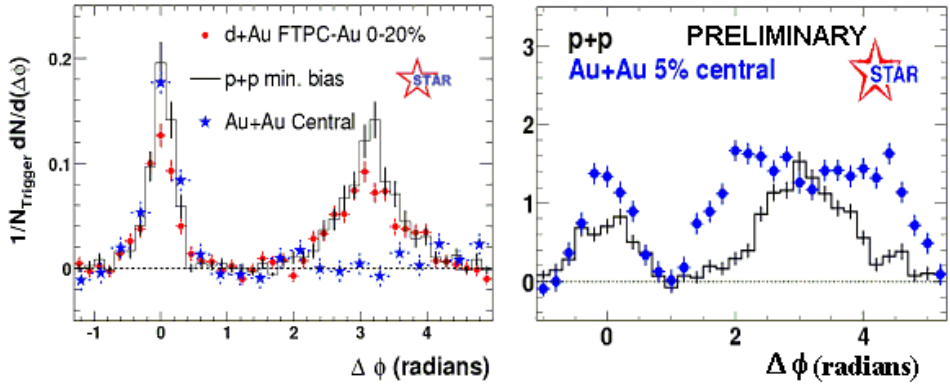,width=0.8\textwidth}
\vspace*{-0.1in}
\caption{Background subtracted azimuth angular correlation between a leading particle of $4$$<$$p_{T}^{\rm trig}$$<$6~GeV/$c$ and an associated particle of 2$<$$p_{T}$$<$$p_{T}^{\rm trig}$ (left panel) and of 0$<$$p_{T}$$<$4~GeV/$c$ (right panel, preliminary) measured by the STAR experiment. For large associated particle $p_{T}$ (left panel), the minimum bias p+p results (histogram) and the $20\%$ central d-Au results (red circles) are similar, while the central Au+Au results (blue stars) are depleted on the away side.  For all associated particle $p_{T}$ (right panel, dominated by low $p_{T}$ particles), the central Au-Au results (blue circles) are enhanced with respect to minimum bias p-p results (histogram).}
\label{FWang_fig1}
\end{center}
\end{figure}

High $p_{T}$ particles are predominantly from jets emerging from initial hard-scatterings between partons. They need finite time to escape the collision zone where a dense medium is formed. The partons and/or fragmented hadrons are expected to lose energy via interactions with the medium, resulting in the so-called jet quenching phenomenon -- suppression of inclusive yield and angular correlation strength at large $p_{T}$. All four RHIC experiments~\cite{fwang} have observed suppressed yield at large $p_{T}$ in central Au-Au collisions. STAR has also observed a suppressed strength of the jet-like angular correlation at large $p_{T}$ as shown in the left panel of Fig.~\ref{FWang_fig1}. It is possible, however, for the suppression to result from initial state effects, such as saturation of gluon density in the Au nucleus. In order to test this possibility, d-Au collisions were measured and found to be similar to p+p, confirming the presumption that jet quenching in central Au+Au collisions is due to final state interactions. 

The depleted jet energy at large $p_{T}$ must be stored in low $p_{T}$ particles. Reconstruction of these low $p_{T}$ particles will therefore serve as a strong experimental confirmation of jet quenching. Moreover, as pointed out by recent theoretical investigations, complete reconstruction of jets and study of their modifications in nuclear medium will provide important insight into the properties of the medium. By studying the amount of energy loss and how the energy is distributed, one may experimentally learn about the medium density, the underlying energy loss mechanism(s), and the degree of equilibration between the energy and the medium. STAR has now performed a complete reconstruction of jets of charged hadrons statistically, by correlating charged hadrons at all $p_{T}$ with a leading particle and subtracting the combinatoric background. The result is shown in the right panel of Fig.~\ref{FWang_fig1}. It was found that the leading particles of the same average $p_{T}$ come from larger energy jets in central Au-Au collisions than in p-p. It was also found that the low $p_{T}$ hadrons are there on the away side of the leading particle and there are more of them in central Au+Au collisions than in p+p. It is observed that, while the near side particle spectral shape is similar in p-p and Au-Au, the away side associated particles are significantly softened in central Au-Au collisions, approaching those from the bulk. The lost energy at high $p_{T}$ has been moved to low $p_{T}$.

Heavy flavor particles play an important role in studying physics of heavy ion collisions. The suppression of heavy quarkonium production is predicted as one of the characteristics of a potential phase transition of nuclear matter from confined to deconfined QGP phase. However, other competing nuclear effects such as parton shadowing, heavy quark energy loss, color screening, and charm recombination will also affect the overall charmonium production in Au-Au collsions. The latest results of particle production in large forward and backward rapidity in d-Au collisions from PHENIX indicate the nuclear effects could play a significant role in interpreting AuAu results. Thus it is very important to systematically measure open charm and $J/\Psi$ production in p-p, d-Au and Au-Au collisions to fully understand the underlying physics.

The latest results of open charm and $J/\Psi$ productions in p+p, d-Au and Au-Au from the  PHENIX and STAR experiments at $\sqrt{s_{NN}}=200$ GeV were presented~\cite{liu}, and the implications of heavy flavor production in cold (d-Au collsions) and hot (Au-Au) nuclear media were discussed.
The ratio of the $J/\Psi$ yields observed in d-Au collsions relative to p-p system, $R_{d-Au}$, was measured to be close to unity at central rapidity $y = 0$, whereas a slight enhancement and suppresion has been observed in backward and forward rapidities, respectively, in d-Au collisions, indicating a possible gluon (anti)shadowing in the forward(backward) directions as well as a possible Cronin kind enhancement in the backward rapidity.



The NA49 collaboration~\cite{botje} has recently completed the
energy scan program at the SPS to search for the onset of a QGP phase
transition. In the energy range of 20--158~$A$~GeV rapid changes occur
in the hadron production properties of central Pb-Pb collisions. They seem to observe a deconfinement phase transition though a purely hadronic
scenario cannot at present be ruled-out. The current hadronic
models are not able to describe the data.

The NA50 experiment~\cite{sitta} measured the $\Psi'$ production in Pb-Pb collisions. In comparison with lighter systems the $\Psi'$ is much more suppressed in nucleus-nucleus than in proton-nucleus interactions, and the suppression pattern is the same in S-U and Pb-Pb as a function of the mean $c{\bar c}$ path in nuclear matter $L$. The anomalous suppression for the $\Psi'$ sets in for lower $L$ values than for the $J/\Psi$.

The NA57 collaboration has reported~\cite{virgili} an enhanced production of $\Lambda$, $\bar\Lambda$, $\Xi$\ and $\bar\Xi$\ when going from p-Be to Pb-Pb collisions at 40 $A$\ GeV/$c$. The enhancement pattern follows the same hierarchy with the strangeness content as at 158 GeV/$c$: $ E(\Lambda) < E(\Xi)$, $ E(\bar\Lambda) < E(\bar\Xi)$. For central collisions the enhancement is larger at 40 GeV/$c$. In Pb-Pb collisions the hyperon yields increase faster at 40 than at 158 $A$\ GeV/$c$. The measurements of the charged particle multiplicity also indicate that values of $dN_{ch}/d\eta$ at the maximum are close to a logarithmic scaling with the center of mass energy.

The NA60 experiment~\cite{arnaldi} reported on the preliminary results from the data collected with Indium-Indium collisions at 158 GeV in October-November 2004. The low mass dimuons have been measured with very high statistics ($\sim10^6$ signal events), good mass resolution ($\sim20-25$ MeV at the $\omega$, $\phi$ masses) and signal to background ratio ($\sim$0.7). The phase space coverage down to zero $p_T$ is reached and full information on associated track multiplicity is available. The integrated $J/\Psi$ suppression in Indium-Indium has been extracted. The ratio between the $J/\Psi$ and the Drell--Yan was shown as a function of the L variable (average length of nuclear matter traversed by the charmonium state after its formation), together with the results of previous experiments (NA38, NA50).

The ALICE Collaboration~\cite{dainese} presented the latest results on 
the expected performance for quenching measurements of charm particles
in Pb--Pb collisions at the LHC.
Due to the large values of their masses, the charm and beauty quarks are
qualitatively different probes of QCD matter formed in heavy-ion collisions, 
since their medium-induced energy loss is predicted to be smaller than that
of massless partons.
Therefore, a comparative study of the attenuation of massless and massive 
probes at the LHC will allow testing the consistency of the interpretation 
of quenching effects as due to energy loss in a deconfined medium and to
further investigate the properties (density) of such a medium.

\begin{figure}[hbt]
  \begin{center}
  \includegraphics[width=0.4\textwidth]{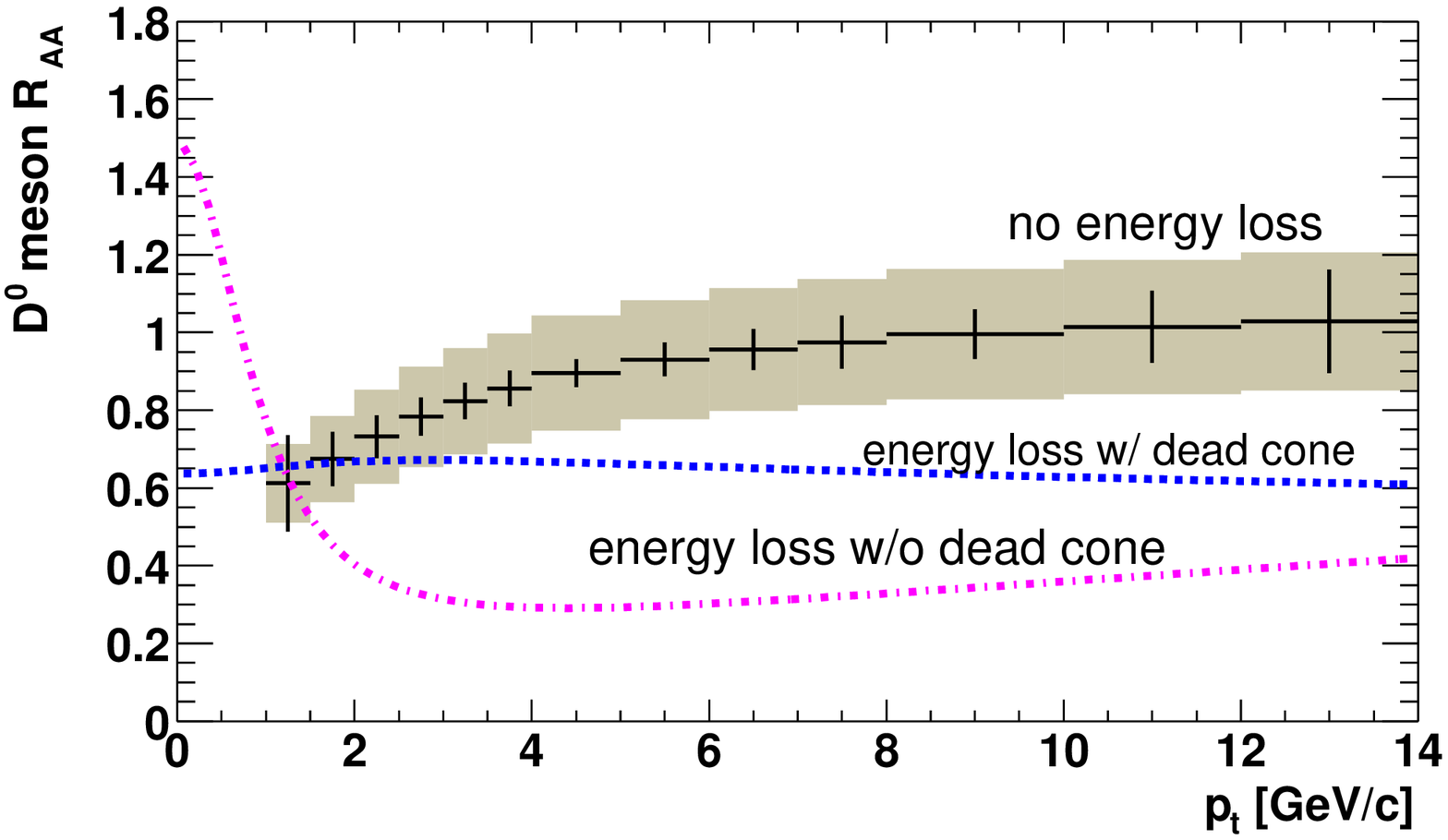}
  \includegraphics[width=0.4\textwidth]{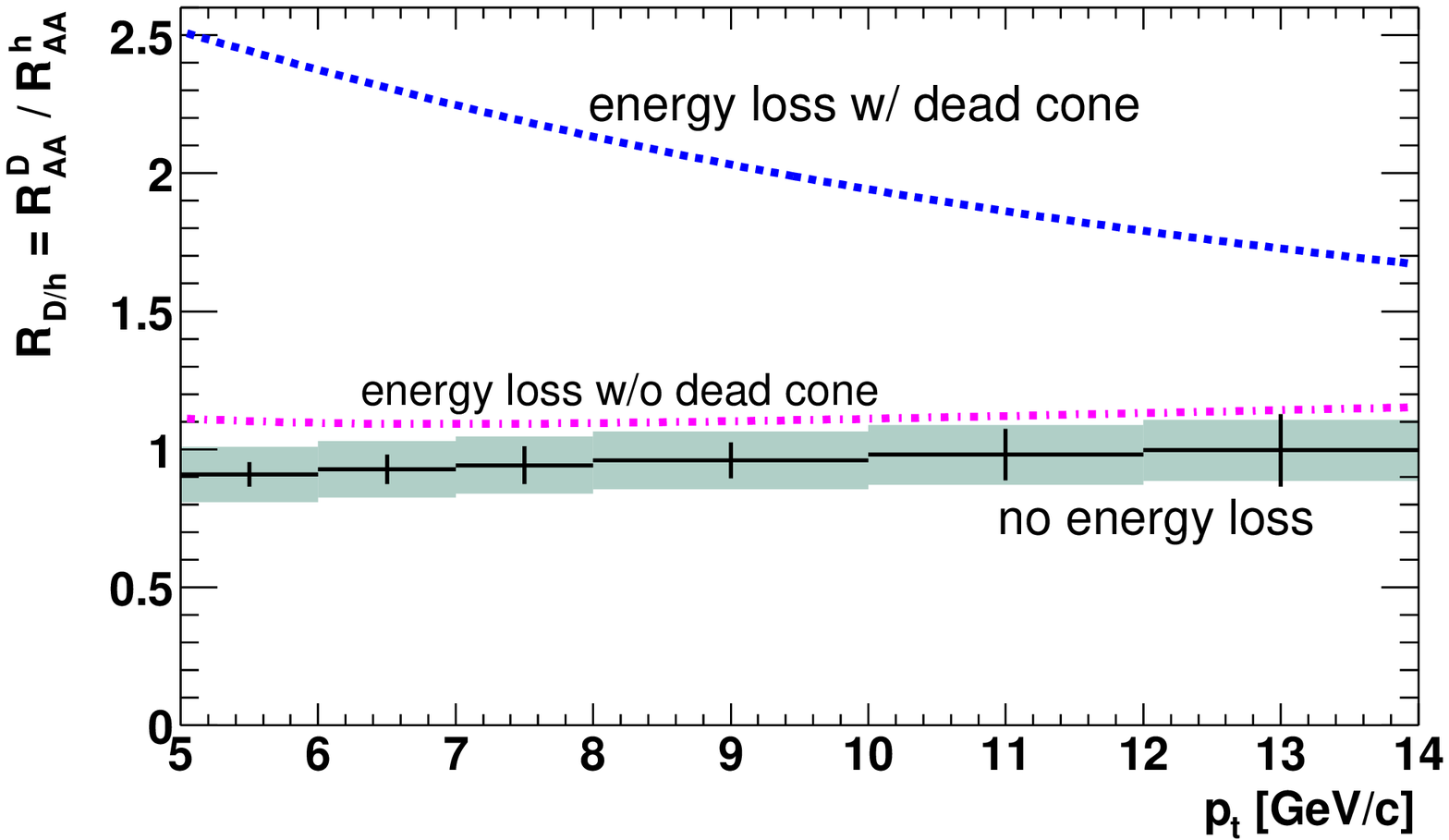}
  \caption{Nuclear modification factor for $D^0$ mesons (left) and
           ratio of the nuclear modification factors for $D^0$ mesons
           and for charged hadrons (right) as expected by ALICE in            1 year of data-taking at LHC luminosity. 
  \label{fig:alice}}
  \end{center}
\end{figure}

It was shown that the exclusive reconstruction of $D^0 \to K^-\pi^+$ decays 
with ALICE allows measurement of the nuclear modification factor\cite{denterria} 
of $D$ mesons and of the $D$/charged hadrons ratio up to 
$p_t\simeq 15~{\rm GeV}/c$. A state-of-the-art model for parton energy loss
was included in the simulation, with a correction for the case of heavy quarks.
The results are presented in Fig.~\ref{fig:alice}. The $D$/charged hadrons 
ratio is found to be enhanced by the effect of the 
charm mass on parton energy loss and, consequently, appears as a 
very clean tool to investigate and quantify this prediction of QCD. 

\section{Core QCD}
\label{sec:qcd}

\begin{figure}[htb]
 \begin{center}
  \epsfig{figure=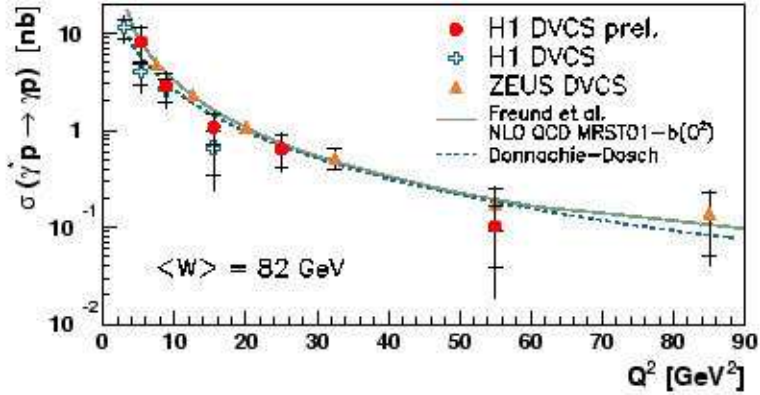,height=6cm}
  \caption{The $\gamma^* p \rightarrow \gamma p$ DVCS cross section as
a function of $Q^2$ for $<W>=82$~GeV with data from H1 and ZEUS
collaborations. The measurement is compared with a NLO QCD prediction
using a GPD parametrization.}
  \label{fig3}
 \end{center}
\end{figure}

A measurement of the inclusive diffractive structure function in deep inelastic scattering extending to the largest $Q^2$ values accessed to date (to $1600$~GeV$^2$) has been reported~\cite{schoeffel} by the H1 experiment in $e^+p$ interactions at {$\sqrt{s}=319$~GeV}. These measurements are well described by the extension of a DGLAP QCD fit realised on previous data at lower $Q^2$ (below $120$~GeV$^2$), which clearly illustrates the hard nature of diffractive phenomena for this kinematic domain. Vector meson production with a hard scale are well dedicated reactions to study the hard nature of diffractive processes. In particular, the diffractive photoproduction $J/\psi$ cross-section has been measured by H1 and ZEUS collaborations to $|t| \le 30$ GeV$^2$. The energy dependance has been shown to be reasonably well described by the BFKL model over the all range of measurements, whereas the DGLAP model describes the data only for $|t| < 5$ GeV$^2$. Another very interesting exclusive process has been studied by both collaborations with the measurement of the Deeply Virtual Compton Scattering (DVCS) cross-section, i.e. the production of real photons. The perturbative QCD calculations for this reaction assume that the exchange involves two partons, having different longitudinal and transverse momenta, in a colorless configuration. These unequal momenta are a consequence of the mass difference between the incoming virtual photon and the outgoing real photon. Therefore, DVCS process gives access to the Generalised Partons Distributions (GPDs), which are generalisations of the familiar parton distributions and incorporate information on correlations between the momenta of partons in the proton. Data are in nice agreement with models beased on GPDs as illustrated in Fig. \ref{fig3}.

The \babar\ Collaboration~\cite{anulli} reported on the studies of the 
processes with hard photon emitted from initial state (ISR). Final states are fully reconstructed and tagged by detection of the hard photon. The invariant mass of the final state determines the effective collision center of mass energy and can be compared to the cross secion of the direct annihilation process. The ISR physics program in \babar consists mainly of light hadron spectroscopy and measurement of the ratio, which provides the experimental input to dispersion integrals for computation of the hadronic contribution to the theoretical estimation of fundamental quantities such as the muon magnetic moment anomaly and the running of the electromagnetic coupling to the $Z$ pole. Preliminary results based on an integrated luminosity of 89.3 $\ifb$, collected at PEP-II, for the final states with four charged hadrons have been presented.
\babar data are in good agreement with previous available results. Moreover, \babar is the only experiment which covers the entire energy range, with an accuracy comparable to the latest precise results from CMD-2 and SND below 1.4 GeV, and much better accuracy of older results from DCI and ADONE above 1.4 GeV. The hadronic contribution for this particular channel evaluated using all available data in the 0.56-1.8 GeV energy range is $a_{\scr \mu}^{\scr hadr} = (14.21\pm 0.87_{\scr exp}\pm 0.23{\scr rad}) 10^{\scr -10}$, while the $\tau$
data give $a_{\scr \mu}^{\scr hadr} = (12.35\pm 0.96_{\scr exp}\pm 0.40_{\scr SU2}) 10^{\scr -10}$. 
The \babar\ data in the same energy region give instead $a_{\scr \mu}^{\scr hadr} = (12.95\pm 0.64_{\scr exp}\pm 0.13{\scr rad}) 10^{\scr -10}$, 
leading to a substantial improvement.


Recent QCD results from LEP were reviewed~\cite{wengler}. The
emphasis is on new studies and on puzzling
disagreements of theory and experiment. The new studies discussed in
more detail are the most precise measurement of unbiased gluon jets
to date, strong evidence of color coherence in 3-jet events, and an,
albeit unsuccessful, search for penta-quarks. As yet unexplained
disagreements are observed in photon-photon collisions for high
momentum charged particle and single jet production, and for the
total cross section of b-quark production.

\begin{figure}[hbt]
\begin{center}
\includegraphics[width=0.3\textwidth]{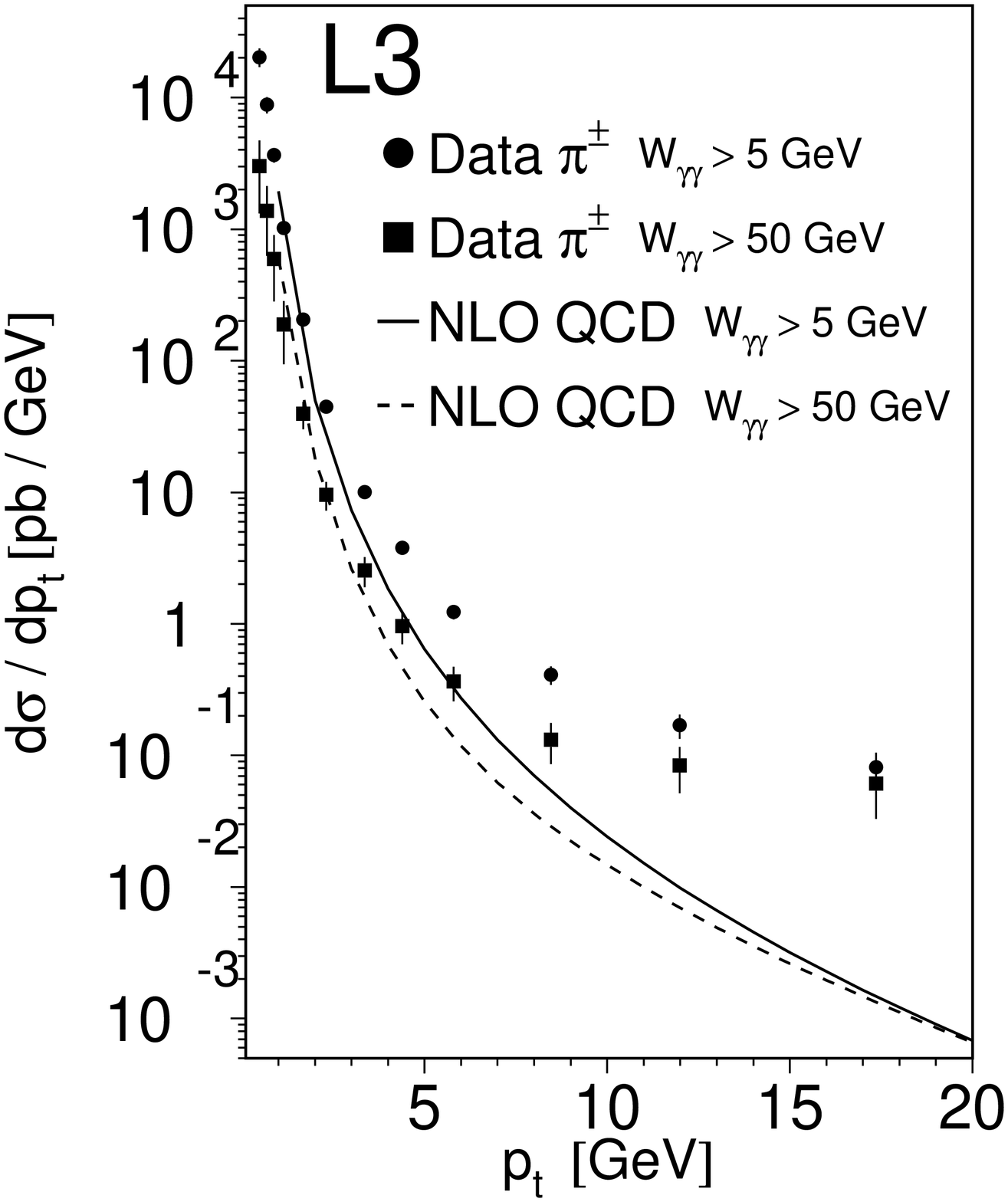}
\includegraphics[width=0.3\textwidth]{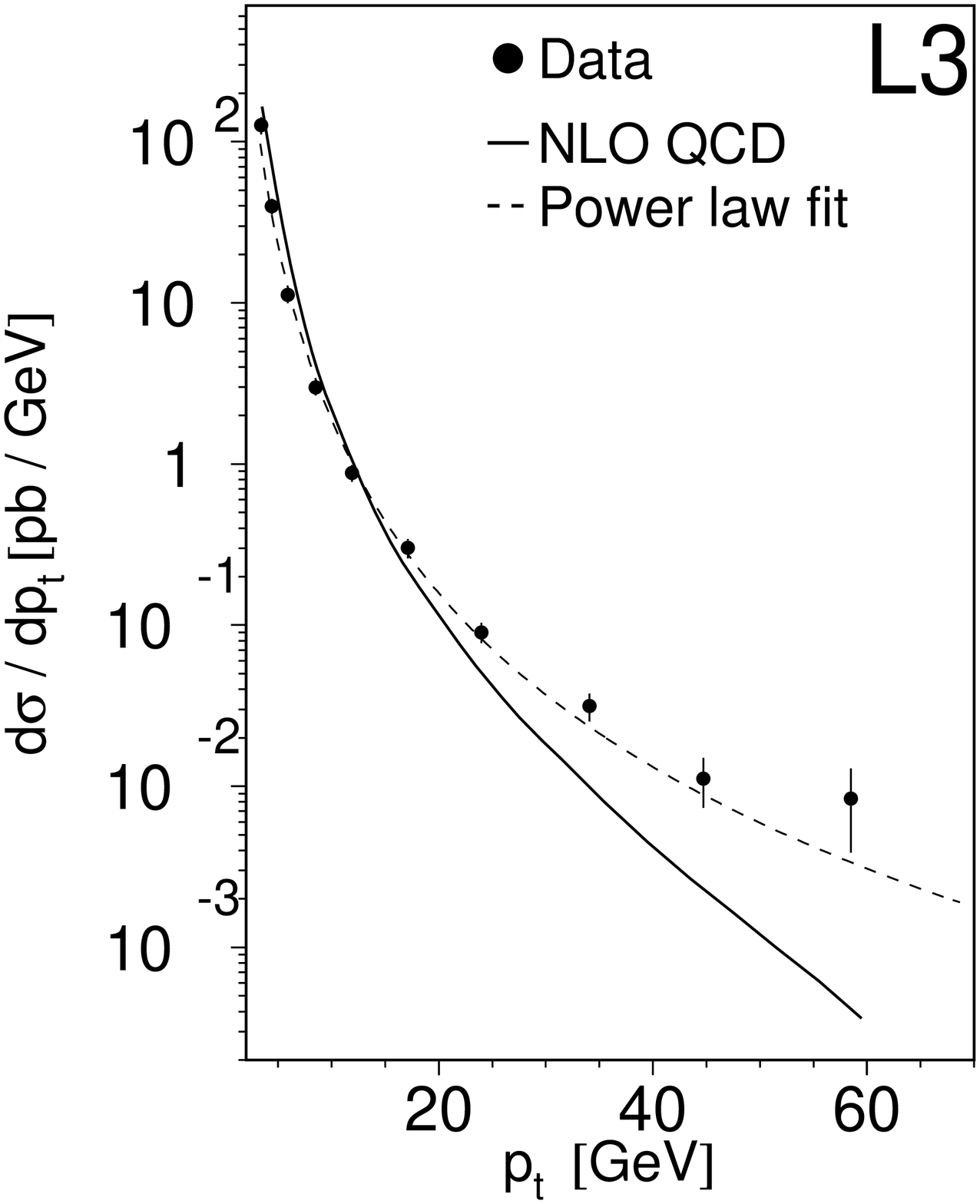}
\caption{The transverse momentum spectra of charged pions for two values of the invariant mass $W$ of the hadronic system (left) and the transverse momentum spectrum of single jets (right) as observed by L3.} 
\label{fig:jetsparticles}
\end{center}
\end{figure}

QCD has been very successful in recent years in describing high
energy hadronic processes such as jet production in $e^+e^-$, $ep$,
and $p\bar{p}$ reactions. It is all the more important to pinpoint
and study areas where discrepancies persist, for
example in the hadronic interactions of two photons, as studied at
LEP2. L3 has studied the production of charged and neutral pions up
to the highest $e^+e^-$ center-of-mass energies available at
LEP2. The left plot in
Figure~\ref{fig:jetsparticles} shows the transverse momentum
spectrum of charged pions for two values of $W$, the hadronic
invariant mass of the photon-photon system. In each case it is
evident that the corresponding calculation in next-to-leading order
(NLO) QCD fails to describe the data for momenta larger than about
4\,GeV. At the highest charged particle momenta measured the theory
underestimates the data by more than an order of magnitude. A
similar measurement by L3 of neutral pion
production leads to the same conclusions. Yet the
presence of high momentum particles should indicate the presence of
a hard scale, and the perturbative calculation should be reliable.
The discrepancy can therefore not easily be understood in terms of
the NLO calculation on parton level. Furthermore at high momenta the
interaction of two photons is expected to be dominated by the
so-called direct process, i.e. the exchange of a fermion, such that
uncertainties in the knowledge of the photon structure are not
expected to be very important. To compare the parton level
calculation to the data, it is folded with the appropriate
fragmentation function. While it is not expected that fragmentation
effects could explain discrepancies of this magnitude, it is
interesting to study the same process by different means. L3 has
also measured the production of single jets in photon-photon
collisions. The right plot of
Figure~\ref{fig:jetsparticles} shows the transverse momentum
spectrum of the jets, again compared to a calculation at NLO QCD. As
in case of the particle spectra, the theory underestimates the data
significantly at high transverse momenta. One has to conclude that
these discrepancies are very significant, and at present not
understood.

\begin{figure}[hbt]
\begin{center}
\includegraphics[width=0.35\textwidth]{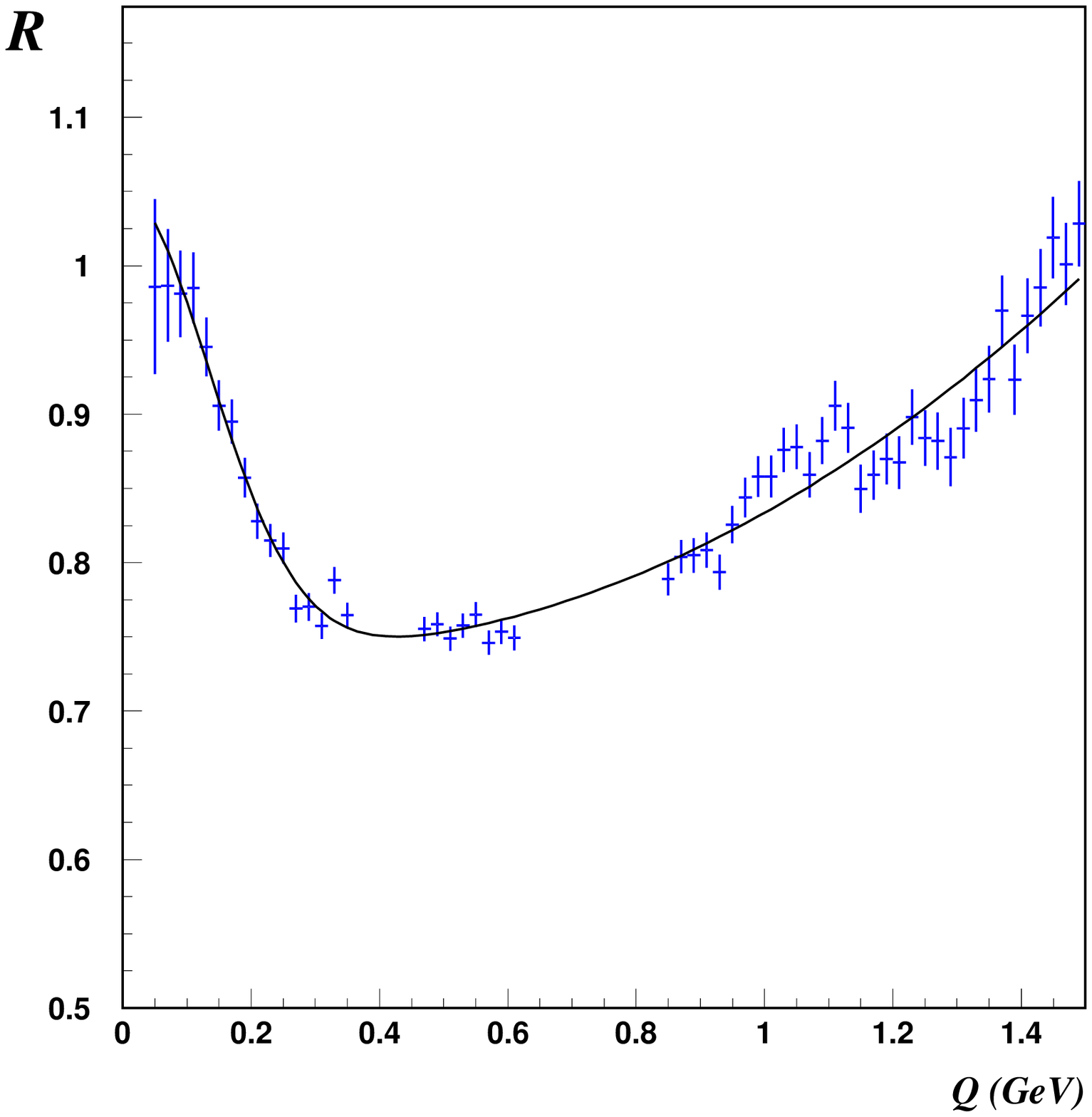}
\includegraphics[width=0.35\textwidth]{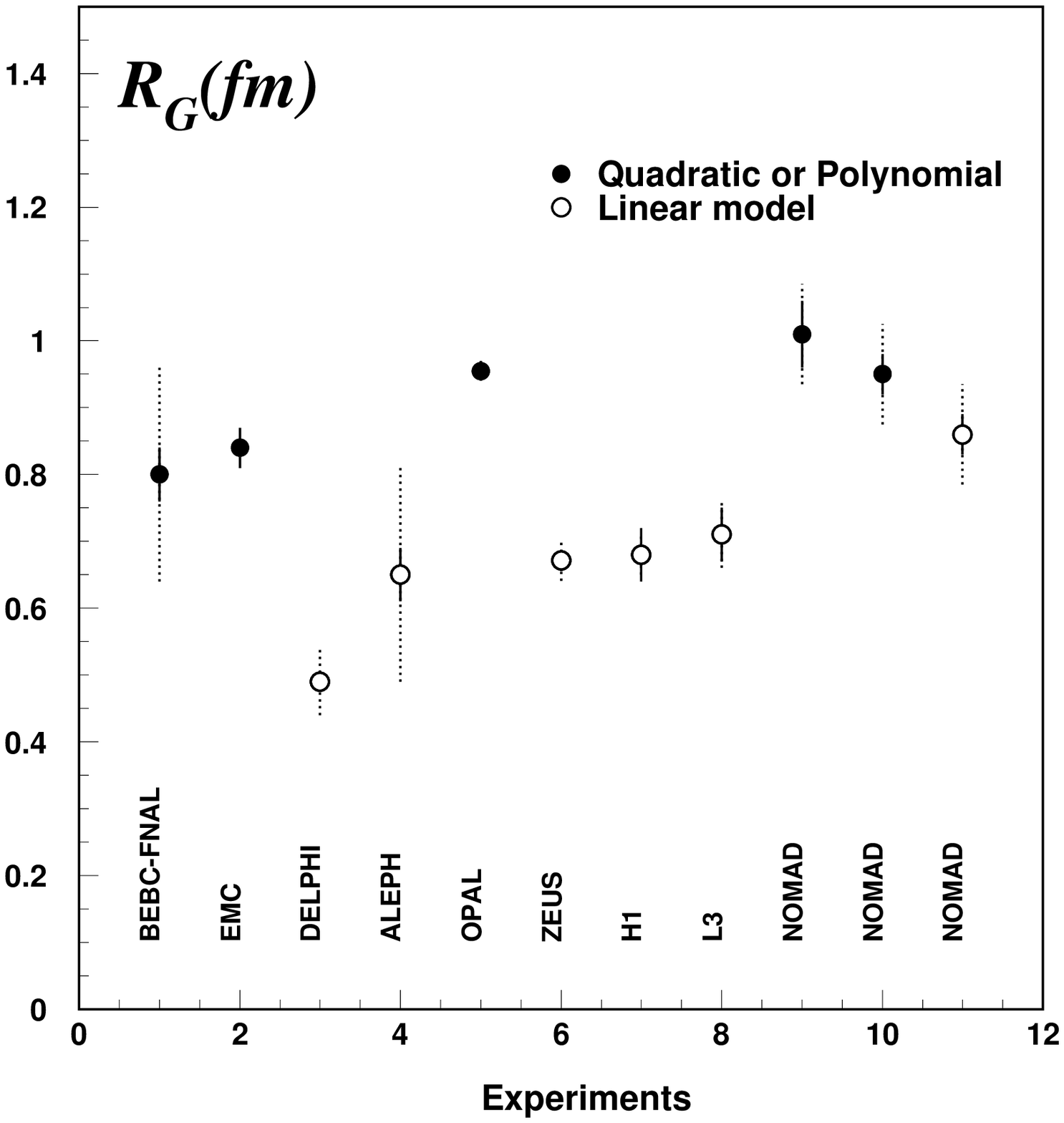}
\caption{BEC effect in NOMAD (left) and compilation of results for the source radius (right).}
\label{fig:Kcomp}
\end{center}
\end{figure}

The NOMAD Collaboration~\cite{zei} reported on Bose-Einstein Correlations 
(BEC) in one and two dimensions in charged current muon-neutrino interactions. In one dimension the Bose-Einstein effect has been 
analyzed with different
parametrizations. The BEC 
parameters are independent of the particle pair charge and of 
the final state rapidity sign of the emitted pions. The two-dimensional shape of the source has been investigated in the longitudinal co-moving frame. A 
significant difference between the transverse ($R_{\bot} = 0.98 \pm 0.10$ fm) and the longitudinal ($R_{||} = 1.32 \pm 0.14$ fm) dimensions has been observed: $(R_{||} - R_{\bot})/R_{||} \approx 35\%$. A weak dependence of both radius and chaoticity on multiplicity and hadronic energy is also found. Performing a 
comparison (see Fig.~\ref{fig:Kcomp}) 
with various results in the $\pi \pi$ channel for lepton-induced reactions yields that the final state hadronization processes have universal features with little dependence on the type or energy of the interacting particles.

The HERMES collaboration~\cite{elschenbroich} presented the first measurements of single--spin asymmetries in semi--inclusive deep inelastic scattering on a transversely polarised hydrogen target. They extracted the asymmetry moments of different sine--dependencies for the production of $\pi^+$, $\pi^-$, and $\pi^0$ mesons. The moment $A_\mathrm{UT}^{\sin(\phi+\phi_S)}$ (see Fig.~\ref{fig:asyms}) can be related to the product of the \textit{transversity} quark
distribution function and the \textit{Collins} fragmentation function.
The transversity is the third distribution function which is required for
a complete description of the quark distributions in the proton at leading
twist. Contrary to the other two distribution functions it is chiral--odd.
Therefore it can be measured in hard
scattering processes only in combination with another chiral--odd object
like the Collins fragmentation function.

\begin{wrapfigure}{r}{6cm}
  \centerline{\psfig{figure=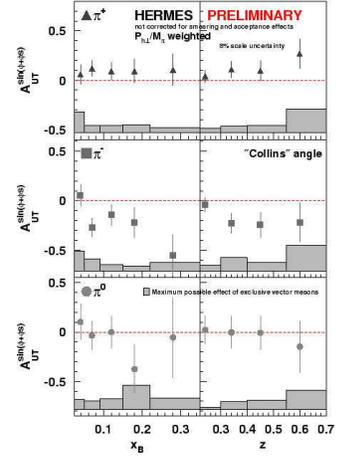,height=6cm}}
  \caption{Asymmetry moments $A_\mathrm{UT}^{\sin(\phi+\phi_S)}$ for
    $\pi^+$, $\pi^-$, and $\pi^0$.
    \label{fig:asyms}}
\end{wrapfigure}

The DELPHI and ALEPH experiments have released~\cite{kucharczyk} new results of the Fermi-Dirac correlations for $p p$ ($\bar{p} \bar{p}$) and Bose-Einstein correlations for $K^{0}_{s} K^{0}_{s}$. Both experiments measure a very small source radius for protons $R \sim 0.1$ $fm$. The source dimension for $K^{0}_{s} K^{0}_{s}$ is in agreement with the previous measurements for $K^{+} K^{-}$, $K^{0}_{s} K^{0}_{s}$. These results, together with earlier LEP measurements, establish the dependence of the correlation radius on the hadron mass.

The recent precision measurement of the muon anomaly $a_\mu$ 
at the Brookhaven National Laboratory
has renewed interest in accurate measurements of the cross
section for $e^+e^-$ annihilation into hadrons. The KLOE experiment has
measured~\cite{denig} the cross section for the Initial State Radiation (ISR) process
$e^+e^-\rightarrow e^+e^-\gamma_{ISR}\rightarrow \pi^+\pi^- \gamma_{ISR}$
at the electron-positron collider DA$\Phi$NE at an energy $W=M_\phi=1.02$ GeV. 
From the dependence of the cross section on the invariant mass of the 
two-pion system $s_\pi$ they extract $\sigma(e^+e^-\rightarrow \pi^+\pi^-)$.
From the cross section measurement they are able to calculate 
the hadronic contribution to the muon anomaly with a negligible statistical
error and with a systematical error of $1.2\% {\rm (exp)} \oplus
1.0\%{\rm (theo)}$.
The resulting value, which is in good agreement with CMD-2's, therefore confirms a $\approx 3 \sigma$ deviation between the direct measurement of the
anomalous magnetic moment of the muon and its theoretical 
calculation. Moreover, it confirms the discrepancy seen between $e^+e^-$ data
(KLOE, CMD-2) and hadronic tau-decay data (ALEPH, OPAL, CLEO), 
which can be used to extract the pion form factor through the CVC theorem, 
taking into account isospin breaking effects. 

Preliminary results from the HERA-B experiment on differential \ptt\ spectra, $A$-dependence and anti-particle to particle ratios were also presented~\cite{zavertyaev}. The anti-particle to particle ratios at HERA-B were measured for $V^0$s and cascade hyperons and the result are shown in Fig.\,\ref{fig:ratio}. Results of the heavy ion experiments studies in Pb-Pb collisions are shown for comparison in the same plot. The \antilambda /\lambdazero\ ratio at HERA-B energy 
is about $20\%$ smaller then at RHIC. At HERA-B energy
this ratio is a rising function and at least a part of the $20\%$ difference may be attributed to the energy dependency and not exclusively to the difference
 between $pA\ -\ AA$ interactions.

\begin{figure}[htb] 
\centering
\includegraphics[width=11.cm]{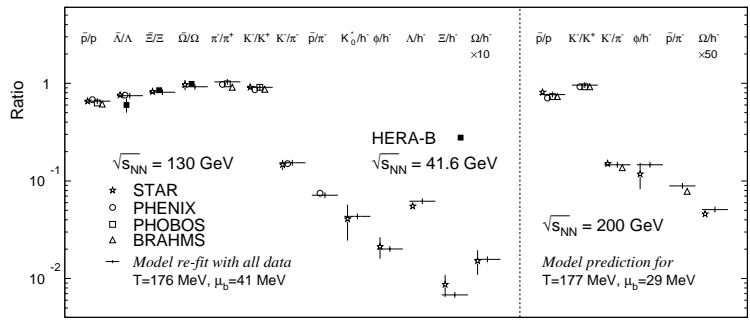}
\vspace{-1.cm}
\caption{Comparison of antiparticle to particle ratio at RHIC and HERA-B.}
\label{fig:ratio}
\end{figure} 

Hadronic jet production with large transverse momentum ($p_T$) provides useful tests of perturbative QCD (pQCD) calculations. The inclusive jet and dijet cross sections at large $p_T$ or large invariant dijet mass ($M_{jj}$) are directly sensitive to the strong coupling constant ($\alpha_s$) and parton density functions (PDF). Deviations from the theoretical predictions at high $p_T$ or $M_{jj}$, not explained by PDF or $\alpha_s$, may indicate physics beyond the Standard Model. CDF has studied~\cite{padley} the multi-jet events by measuring event properties in the ``transverse'' region between the jets.  D\O\ has measured~\cite{padley} angular decorrelations to examine this physics. These measurements become all the more interesting now that the Tevatron has entered a new era of luminosity and energy. The center of mass energy of the Tevatron has risen to $\sqrt{s}=1.96 $TeV from $\sqrt{s}=1.8 $TeV and during the early running the experiments have already accumulated more luminosity that all of Run I.  As a consequence the CDF and D\O\ experiments are now probing length scales on the order of $10^{-19}m$ and the discovery of new physics is a tantalizing possibility.  To give a sense of the physics reach, CDF has seen an event with 1.3 TeV/c$^2$ dijet energy and D\O\ has seen one at 1.2 TeV/c$^2$.

The D\O\ and CDF collaborations also reported preliminary results on the W 
and Z cross sections at 1.96~TeV~\cite{hesketh}.
The electron and muon decay channels provide the cleanest signals. The inclusive production cross sections test our understanding of the 
boson production mechanism, as well as the detector performance and 
calibration.
Differential cross sections with various jet multiplicities and jet 
energies provide a precision test of perturbative QCD.
Other parameters, such as the boson transverse momentum, rapidity and 
forward-backward asymmetries are sensitive to parton distribution functions.

Both D\O\ and CDF measure inclusive cross sections which are in good 
agreement with NNLO theoretical predictions.
Uncertainties vary in the different channels and at the different 
experiments, but are dominated by the luminosity determination (6 - $10\%$), lepton identification efficiencies ($<3.5\%$) and PDF 
uncertainties (1 - $1.5\%$).
Taking the ratio of the W and Z cross sections, many of these systematic 
uncertainties cancel.
Combining the results from the electron and muon channel, CDF obtain a 
new preliminary result of $\mathrm{R} = 10.94 \pm 0.15 (\mathrm{stat}) 
\pm 0.13 (\mathrm{syst})$.
 From this ratio it is also possible to extract an indirect measurement 
of the W width, $\Gamma_W = 2071 \pm 40$~MeV, 
which can be compared to the current world average of $\Gamma_W = 2118 
\pm 42$~MeV and the theoretical prediction of $\Gamma_W = 2092 \pm 2.5$~MeV.

The inclusive di-photon cross-section was measured in the central region and found to be in good agreement
with NLO QCD predictions.  Cross-sections were presented~\cite{mcnulty} for events containing
an energetic photon in addition to a heavy-flavor jet.  The ratio of photon+c
to photon+b events is also measured.  Results are currently statistically limited and in
agreement with Pythia predictions.

\section{Heavy Flavors}
\label{sec:hf}

The main reason that charm physics is attractive at the Tevatron is the very
high production cross-section, allowing the detectors~\cite{campanelli} to record huge 
(in many cases, the world's largest) samples of charm and beauty hadrons.
A very large $D^0$ sample
was recorded by CDF, that can be made very pure by requiring an additional
``slow'' pion from $D^{*+}$ decay, with small $M(D*)-M(D^0)$ difference. CDF has published cross section measurements for $D^0$, $D^+$, $D^{*+}$ and $D_s$
production, which are in agreement with the upper side of the theory prediction. The large $D^0$ sample was also used to measure relative branching ratios and search for CP asymmetries.

 The BES collaboration observed~\cite{yuan} for the first time
$\pspto \kskl$ and gave improved measurement on
$\jpsito \kskl$ branching fraction (see Fig.~\ref{beskskl}). From these results, the relative phase
between the three-gluon and the one-photon annihilation amplitudes
of the $\psp$ decays to pseudoscalar meson pairs is measured to be
either $(-82\pm 29)^{\circ}$ or $(+121\pm
27)^{\circ}$. Comparing $\psp$ and $\jpsito \kskl$
branching fractions, one gets
      \( Q_h = \frac{\BR(\pspto \kskl)}{\BR(\jpsito \kskl)}
                   = (28.8\pm 3.7)\% \),
which indicates that $\psp$ decays is enhanced by more than
4$\sigma$ relative to the ``$12\%$ rule'' expected from pQCD, while
for almost all other channels where the deviations from the ``$12\%$
rule'' are observed, $\psp$ decays are suppressed.

\begin{wrapfigure}{r}{6cm}
  \centerline{\psfig{figure=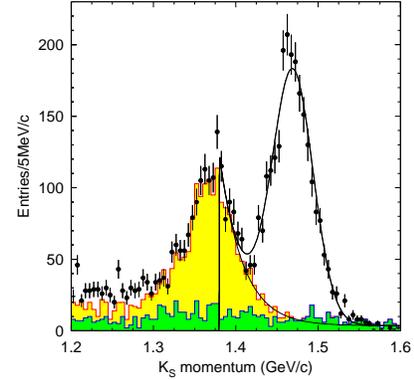,height=5.0cm}}
\vspace*{-0.1in}
  \caption{The $K_S$
momentum distribution at $\jpsi$ for BES data.. The dots with error bars are data and the curves are the
best fit of the data. Dark shaded histogram is from $K_S$ mass
side band events whereas , light shaded one is from backgrounds.
  \label{beskskl}}
\end{wrapfigure}

The BES collaboration also observed for the first time four
$Vector \,\, Tensor$ (VT) decay channels $\pspto \omega
f_{2}(1270)$, $\rho a_2(1320)$,
$K^*(892)^0\overline{K}^*_2(1430)^0+c.c.$ and $\phi
f_2^{\prime}(1525)$, which are suppressed by a factor of 3 to 5
compared with the pQCD ``$12\%$ rule'' expectation; improved the
$\pspto\pi^0\jpsi$, $\eta\jpsi$ and $\gamma\chi_{c1,2}$ decay
branching ratios measurements in $\gamma \gamma \jpsi$, $\jpsito
e^+e^-$ or $\mu^+\mu^-$ final states; measured
the relative branching ratio between $\chicJto \aab$ and
$\chicJto \ppb$ to test the color-octet mechanism
calculation; and a search for the CP violating processes $\jpsito
\ksks$ and $\pspto \ksks$ for testing the EPR paradox.

The CLEO experiment presented~\cite{napolitano} recent results on charm physics from CLEO-III, emphasizing QCD and hadronic structure. These include the decay $\Xi_c^0\to pK^-K^-\pi^+$ and the form factor for $D^0\to\pi^-e^+\nu_e$. They also discuss upcoming measurements with CLEO-c, including dramatic improvements in the  $D^0\to\pi^-e^+\nu_e$ form factor, and glueball spectroscopy in $J/\psi\to\gamma X$. CLEO-c will obtain a very large sample of $e^+e^-\to D^0\bar{D}^0$ events. 


The FOCUS experiment presented~\cite{vaandering} measurements of the properties of the $L=1$ excited $D$
mesons, $D_2^{*0}$ and $D_2^{*+}$, and evidence for broad states interpreted as
the $D_0^{*0}$ and $D_0^{*+}$. The parameters of the 
$D_2^{*0}$ and $D_2^{*+}$ were found to be 
$M_{D_2^{*0}} = 2464.5 \pm 1.1 \pm 1.9 \textrm{~MeV}/c^2$, 
$\Gamma_{D_2^{*0}} = 38.7 \pm 5.3 \pm 2.9 \textrm{~MeV}/c^2$, 
$M_{D_2^{*+}} = 2467.6 \pm 1.5 \pm 0.8 \textrm{~MeV}/c^2$, and 
$\Gamma_{D_2^{*+}} = 34.1 \pm 6.5 \pm 4.2 \textrm{~MeV}/c^2$ respectively. 
The existence of a broad state in the neutral final state agrees with results
from Belle and the evidence for a broad state in the charged final state is a
first observation. If interpreted as $D_0^{*0}$ and $D_0^{*+}$ the parameters
of the broad state excess are found to be 
$M_{D_0^{*0}} = 2407 \pm 21 \pm 35 \textrm{~MeV}/c^2$, 
$\Gamma_{D_0^{*0}} = 240 \pm 55 \pm 59 \textrm{~MeV}/c^2$, 
$M_{D_0^{*+}} = 2403 \pm 14 \pm 35 \textrm{~MeV}/c^2$, and 
$\Gamma_{D_0^{*+}} = 283 \pm 24 \pm 34 \textrm{~MeV}/c^2$ respectively. These
results have been recently published.

\begin{figure}[hbt]
  \begin{center}
  \includegraphics[width=6cm]{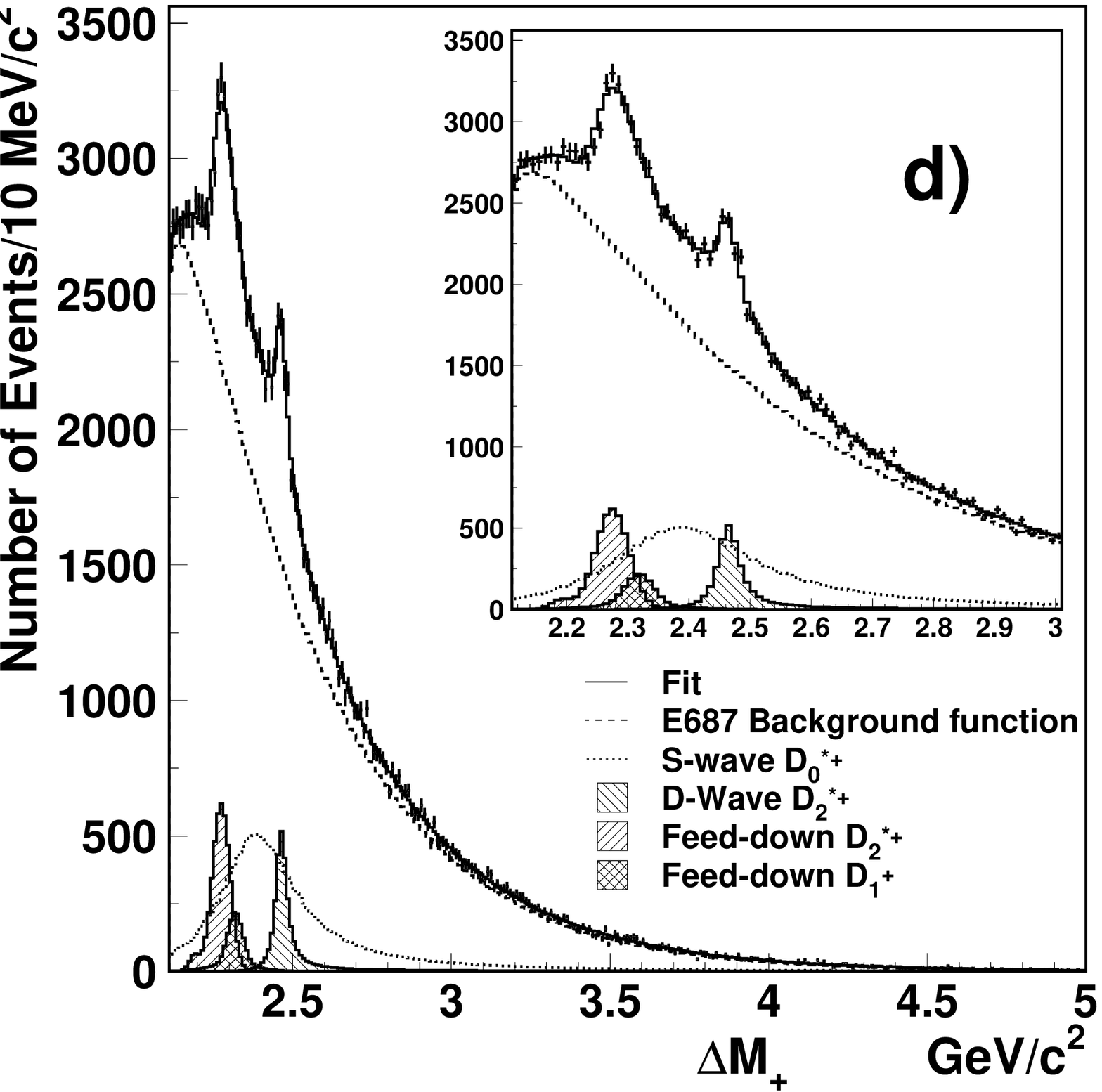}  
  \includegraphics[width=6cm]{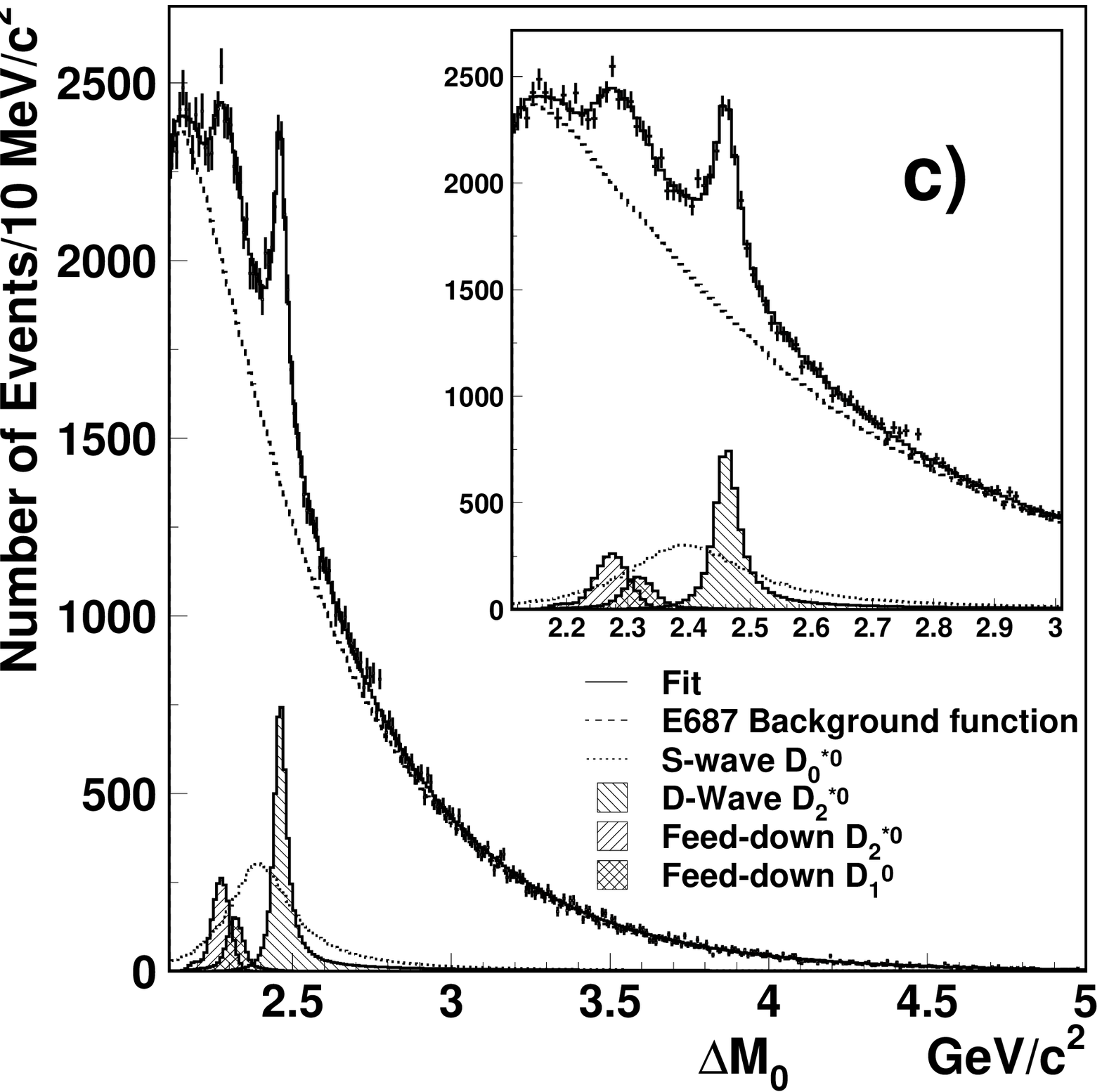}
  \end{center}  
  \caption{Fit of FOCUS signals for $D_2^{*0}$ (left) and  $D_2^{*+}$ (right) and evidence for
  $D_0^{*0}$ and $D_0^{*+}$. The  $D_0^{*0}$  is also seen in a recent report
  from Belle, the $D_0^{*+}$ is the first such observation.}
\end{figure}

The $b\bar{b}$ and $\Upsilon$ production cross section have been measured~\cite{nedden} in collisions of $920~\mbox{GeV}$ protons off a 
nuclei target using the HERA-$B$ detector. Within the invariant mass spectra of $e^{+}e^{-}$ or $\mu^{+}\mu^{-}$
events originating from a $J/\psi$ decay, the $\Upsilon$ search was performed. The most recent measurements, 
using data collected in 2002/2003, yield a cross section in the combined analysis of $\sigma(b\bar{b}) = (12.3^{+3.5}_{-3.2})~\mbox{nb/nucleon}$ and for the hidden beauty production $\sigma(\Upsilon \rightarrow l^{+}l^{-})$ =
$(3.4 \pm 0.8)~\mbox{pb/nucleon}$.

The H1 and ZEUS collaborations reported~\cite{melzer}
on beauty production at the HERA collider. Differential distributions for both
photoproduction and deep inelastic scattering were presented and compared to NLO QCD 
calculations and LO+parton shower Monte Carlo simulations. All recent measurements 
show good agreement between data and NLO QCD calculations. 
The Monte Carlo simulations describe the shape very well, though sometimes the
predictions are too low in normalisation. The higher statistics which will be collected in the HERA II running period in the next years will allow more precise differential measurements.

The CLEO Collaboration reported~\cite{shepherd} inclusive semileptonic $B$ decays.  An improved inclusive branching fraction, $\mathcal{B}\left(B\to Xe^+\nu_e\right) =  \left(10.91\pm0.09\pm0.24\right)\%$, was presented.  Moments of the distributions of lepton energy and recoil hadronic mass were measured as a function of minimum lepton energy requirement.  These moments were used to constrain Heavy Quark Effective Theory parameters $\bar{\Lambda}$ and $\lambda_1$ in the context of the pole-mass scheme.  When the theory parameters are fixed using the 1.5 GeV minimum lepton energy cut and the CLEO $b\to s\gamma$ measurement an increasing discrepancy between theory and experiment emerges in the first moment of the lepton energy spectrum as the cut is lowered.

Highlights of recent CLEO studies on the $\Upsilon (1S)$, $\Upsilon (2S)$, and
$\Upsilon (3S)$ were reported~\cite{duboscq}. The preliminary results include the first observation
 of a hadronic cascade in the $\Upsilon$ system not involving  pions,  $\Upsilon (3S) \to \gamma \chi_{b1,2} (2P)$ followed by    $\chi_{b1,2} (2P) \to \omega \Upsilon (1S)$, a search for inclusive $\psi$ production in $\Upsilon(1S)$ decays, and a search for nine different two  body hadronic decays of the $\Upsilon(nS)$, $n=1,2,3$, resulting in the discovery of two such decays, and vastly improved upper limits on  
 the branching ratios of the rest.

\babar\ reported~\cite{smith} new results for five charmless hadronic $B$ decays.
These included the first limits for decays involving the scalar
$a_0(980)$.  The sector of decays $B\to (\eta,\etapr)(K,\Kstar)$ is now
well measured.  The large branching fraction seen by CLEO for \etaKst\ is
confirmed and $90\%$ CL upper limits on \etapKst\ restrict the size of a
possible color-singlet amplitude.  Searches for 
eight isoscalar $B$ decays as well as charmless modes with \piz\ mesons place restrictive limits on the
departure of the $CP$-violating parameter $S$ measured in \etapKz\ from
the value of $\sin 2\beta$ measured in $B\to J/\psi\KS$.

The most noteworthy new measurement is a full angular analysis of the
decay $B\to\phi\Kstarz$.  The longitudinal polarization in this decay is
found to be $0.52\pm0.07\pm0.02$; this
value is surprising since naive expectations and measurements for
$B\to\rho\rho$ indicate a value very close to 1.  This confirms earlier
measurements by \babar\ and Belle and is still not understood theoretically.

Belle measured~\cite{zhang} the branching fractions of the decays $B\to\phi
K^*$, $B^+\to \rho^+\rho^0$. They observed the decay $B^+\to \rho^+\pi^0$,
and the first evidence for $B^0 \to \rho^0\pi^0$.
An angular analysis is performed on the $VV$ modes.
It indicates that, in the tree-dominated $B\to\rho^+\rho^0$, the SM
prediction $R_0 \gg R_\perp$ is confirmed.
In contrast, in the pure $b\to s$ penguin $B\to\phi K^*$ they find
$R_0\approx R_\perp$; also find $R_\perp\gg R_\parallel$
($R_0+R_\perp+R_\parallel=1$). Both of these results for $B\to\phi
K^*$ are in disagreement with SM predictions.
It is thus important to obtain polarization measurements in
other modes, especially the pure penguin $b\to s \bar d d$ decay,
$B^+\to K^{*0}\rho^+$.

B physics results as well as future prospects were also reported~\cite{yip} from the CDF and D{\O} experiments at the Tevatron Run II at Fermilab.  This includes various B mass and lifetime measurements, B mixing, rare decays, and spectroscopy. CDF has had the best mass measurements for $\rm B_s$ and $\rm \Lambda_b$. {D\O} has collected the world's largest sample of
exclusive $\rm B_s \rightarrow J/\psi + \phi$ decays.  The 
limits on the branching fractions
$\rm Br(B_s \rightarrow \mu^+ \mu^-)$ and
$\rm Br(B_d \rightarrow \mu^+ \mu^-)$ measured by CDF are
also the world's best published ones.

\section{Top and Higgs}
\label{sec:wth}

At the Tevatron, top quarks are mainly produced 
in  pairs ($t\bar{t}$). The measurement of the $t\bar{t}$ production 
cross-section, $\sigma(t\bar{t})$, serves as a test of perturbative QCD
predictions. The two Tevatron experiments, CDF and D\O\ have presented~\cite{ia}
preliminary  results
on  $\sigma(t\bar{t})$ at $\sqrt{s}=1.96$~GeV using Tevatron Run II data.

In the Standard Model (SM) the top quark decays to a
$W$ boson and a $b$-quark almost $100\%$ of the times. When both $W$s decay 
leptonically, the final state contains two high-$p_T$ isolated leptons,
two $b$-jets, and a missing transverse energy ($E_T^{miss}$). Several SM
processes, $Z/\gamma^*\to \ell \ell$, $WW$, $WZ$,
$ZZ$, $W+jets$, with $\ell = e, \mu,$ or $\tau$, contribute to the background.
CDF and  D\O\ 
have both presented results on $\sigma(t\bar{t})$ using 
$t\bar{t}  \to ee / \mu\mu / e\mu + 2 jets + E_T^{miss}$ events.

\begin{wrapfigure}{r}{7cm}
\vspace*{-0.3in}
  \centerline{\psfig{figure=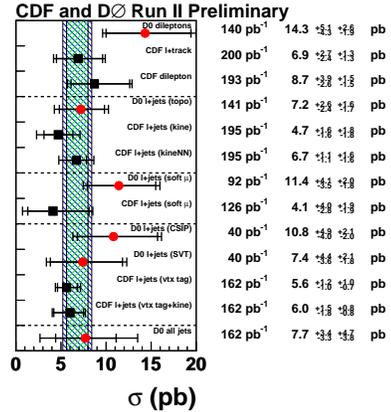,height=6.0cm}}
\vspace*{-0.1in}
\caption{Top pair production cross-section measurements in various final states
at $\sqrt{s}=1.96$~TeV by CDF and D\O\ experiments. The hatched areas correspond to theoretical calculations.
\label{fig:ttbar_xsection}}
\end{wrapfigure}

Top pair events with one W decaying leptonically and the other to hadrons
give rise to $t\bar{t}  \to \ell + \ge 4 jets + E_T^{miss}$ event
topology. Here dominant backgrounds are $W+jets$ and QCD multi-jet
production with a jet faking an isolated lepton.
D\O\ has measured $\sigma(t\bar{t})$ using event 
topological information. Similar approach has been used by CDF in
two independent analyses, one of which combines several kinematic
variables into  a Neural Network and thereby achieves better signal to
background discrimination.

The presence of two b-quarks in the $t\bar{t}$ events 
gives powerful tool for background suppression by tagging
b-jets either via soft muon from $b\to \mu / b\to c \to \mu $ or through 
displaced vertices. Soft muon tagging technique has been used by both, the CDF
and  D\O\ collaborations in 
$t\bar{t}  \to \ell + \ge 4 jets + E_T^{miss}$ final states. CDF has performed
two analyses using b-tagging with the displaced vertices, one of which
makes detailed use of event kinematic shape.  D\O\ has presented  two
analyses 
for $\sigma(t\bar{t})$ with two different techniques of b-jet
lifetime tagging -- Secondary Vertex Tagging, SVT, and Counting Signed
Impact Parameter tagging, CSIP.

Top pair events with both $W$s decaying to hadrons produce events
with 6 or more jets. In this case extraction of the signal from
overwhelming QCD multi-jet background is only feasible by employing
b-tagging techniques. D\O\  has presented measurement of
$\sigma(t\bar{t})$ in multi-jet final state using b-tagging through
displaced vertices.

\begin{wrapfigure}{r}{6cm}
\vspace*{-0.3in}
  \centerline{\psfig{figure=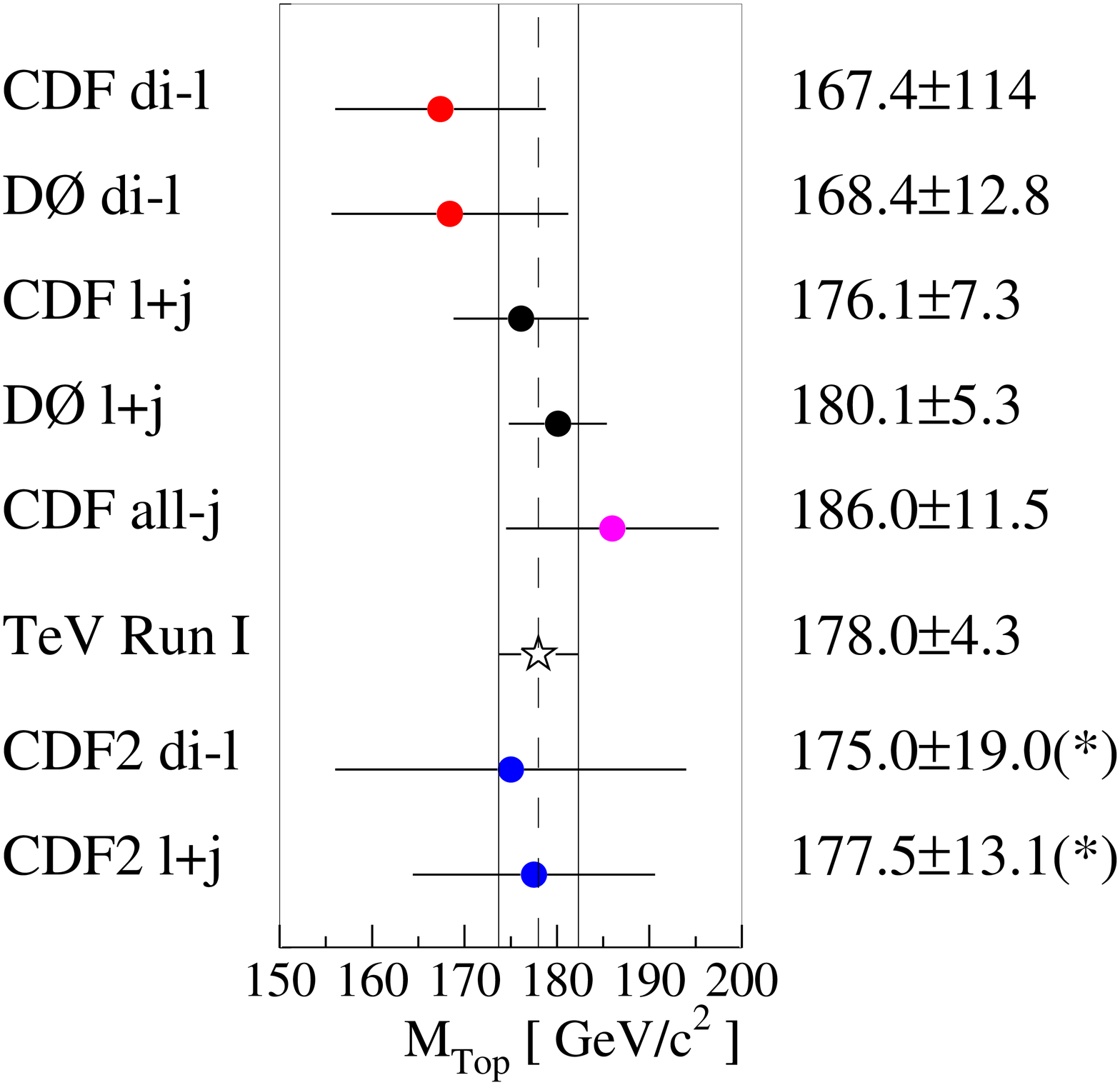,height=5.0cm}}
\vspace*{-0.1in}
\caption{Summary of the measurements of the top quark mass at the Tevatron. (*): Preliminary result.
\label{fig:summary}}
\end{wrapfigure}

Figure~\ref{fig:ttbar_xsection} summarizes the measurements of
$\sigma(t\bar{t})$ in various final states. The displaced hatched bands
indicate  theoretical predictions on the cross-section. 
All measurements, although limited in statistics, are consistent with the predictions.

Preliminary results on the measurement of the top quark mass at the Tevatron Collider were presented~\cite{cerrito}. In the dilepton decay channel, CDF measures $m_t$=175.0$^{+17.4}_{-16.9}$(stat.)$\pm$8.4(syst.) GeV/$c^2$, using a sample of $\sim$126 pb$^{-1}$ of proton-antiproton collision data at $\sqrt{s}=1.96$ TeV (Run II). In the lepton plus jets channel, the CDF Collaboration measures 177.5$^{+12.7}_{-9.4}$(stat.)$\pm$7.1(syst.) GeV/$c^2$, using a sample of $\sim$102 pb$^{-1}$ at $\sqrt{s}=1.96$$~$TeV. The D\O\ Collaboration has newly applied likelihood technique to improve the analysis of $\sim$125 pb$^{-1}$ of proton-antiproton collisions at $\sqrt{s}=1.8$ TeV (Run I), with the result: $m_t$=180.1$\pm$3.6(stat.)$\pm$3.9(syst.) GeV/$c^2$. The latter was combined with all the measurements based on the data collected in Run I to yield the most recent and comprehensive experimental determination of the top quark mass: $m_t=178.0\pm2.7({\rm stat.})\pm3.3({\rm syst.})$ GeV/$c^2$. The new measurements, as well as those which were not updated recently, are shown in Figure \ref{fig:summary}.


Both the CDF and {D\O} collaborations have performed a search for single top quark production in the lepton+jets final state at $\sqrt{s}=1.96$ TeV, using samples of about 160 pb$^{-1}$. In the case of the CDF analysis, a combined (both s- and t-channels) search and a dedicated t-channel search have been carried out. The signal contribution is extracted from a fit to a discriminant variable. No excess above expectation is observed in data. The single top cross-section limits at 95$\%$ CL obtained are $\sigma_{s+t}<13.7$ pb and $\sigma_t<8.5$ pb. The {D\O} analysis, still under optimization, reported expected upper limits at the preselection level already lower than those obtained in Run I after final selection. Significant improvements in both analyses are expected in the near future. The CDF Collaboration has performed a measurement of the ratio B($t \rightarrow Wb$)/B($t \rightarrow Wq$) using a sample of 108 pb$^{-1}$ collected in Run II. This measurement is based on the examination of the $b$-tagging rates in the lepton+jets channel. The most likely value found in data is B($t \rightarrow Wb$)/B($t \rightarrow Wq$)$=0.54^{+0.49}_{-0.39}$, consistent with the SM value of $\simeq 1$.

The {D\O} Collaboration has performed a measurement of the fraction of longitudinally polarized $W$ bosons in top
quark decays ($f_0$), using a sample of 125 pb$^{-1}$ collected in Run I.
This analysis considers the lepton+jets final state and extracts $f_0$ making use of a sophisticated
likelihood fitting procedure, whose previous application to the top quark mass yielded the world's 
single most precise measurement.
The result obtained is $f_0=0.56\pm 0.32$, in good agreement
with the SM expectation of $\simeq 0.7$.

\begin{wrapfigure}{r}{7cm}
\vspace*{-0.5in}
  \centerline{\psfig{figure=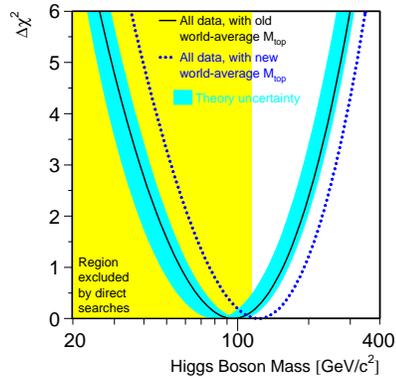,height=6.0cm}}
\vspace*{-0.1in}
\caption{Standard Model fit using the new combined Top mass from CDF and {D\O} compared with the previous fit.
\label{fig:sm_fit}}
\end{wrapfigure}

The Standard Model Fit~\cite{roth} shows good agreement between measurements and theory prediction. The measurments are internally consistent. The overall fit prefers a light Higgs mass. The new combined Top mass from CDF and {D\O} shifts the central value of the Higgs mass fit to 117 GeV.  The new upper limt ($95\%$ C.L.) of the Higgs mass is now 251 GeV. Figure~\ref{fig:sm_fit} shows the result of the new fit compared with the older one.

Both collider experiments, CDF and D\O\ have presented
the SM Higgs searches based on the data collected in Run~II~\cite{avto}.
At low Higgs masses, $m_H \lsim 135$ GeV, the $WH$ and $ZH$ associated
productions can be explored with vector bosons decaying into leptons. Since the
production cross sections are very small, on the order of 0.1~pb or less,
the integrated luminosities in excess of a few fb$^{-1}$ are needed
to discover, or rule out, the SM Higgs. However, detailed
studies of the background processes have begun, and new limits
on the Higgs boson production have been presented by both experiments.

The D\O\ collaboration has analysed the $W \rightarrow e\nu$ decays accompanied
by two b-tagged jets, and has obtained good agreement between data
and expectations from MC. Based on 174~pb$^{-1}$ of data
an upper limit on the $Wb\bar{b}$ production cross section,
$\sigma[Wb\bar{b}] < 20.3$~pb has been set at $95\%$ C.L. A limit
has also been obtained for the $WH$ associated production,
$\sigma[W(\rightarrow e \nu)H(\rightarrow b\bar{b})]~<~12.4$~pb for
a Higgs masss of 115 GeV. Using electron and muon decays of the $W$,
CDF has set an even better limit of $\lsim$~5~pb depending on the Higgs 
mass. This result is based on
an integrated luminosity of 162~pb$^{-1}$ and is shown in
Fig.~\ref{fig:cdfd0_smhiggs}a.

At larger masses, $m_H \gsim 135$ GeV, the
$H\rightarrow WW\rightarrow\ell\nu\ell\nu \; (\ell = e, \mu)$, final
states are promising to look at. D\O\ 
has analysed 150~-~180~pb$^{-1}$ of data depending on the final
state, and limits on the Higgs boson production cross section
times branching
fraction into $W$ bosons are shown in Fig.~\ref{fig:cdfd0_smhiggs}b.

\begin{figure}[hbt]
\centerline{
\epsfig{figure=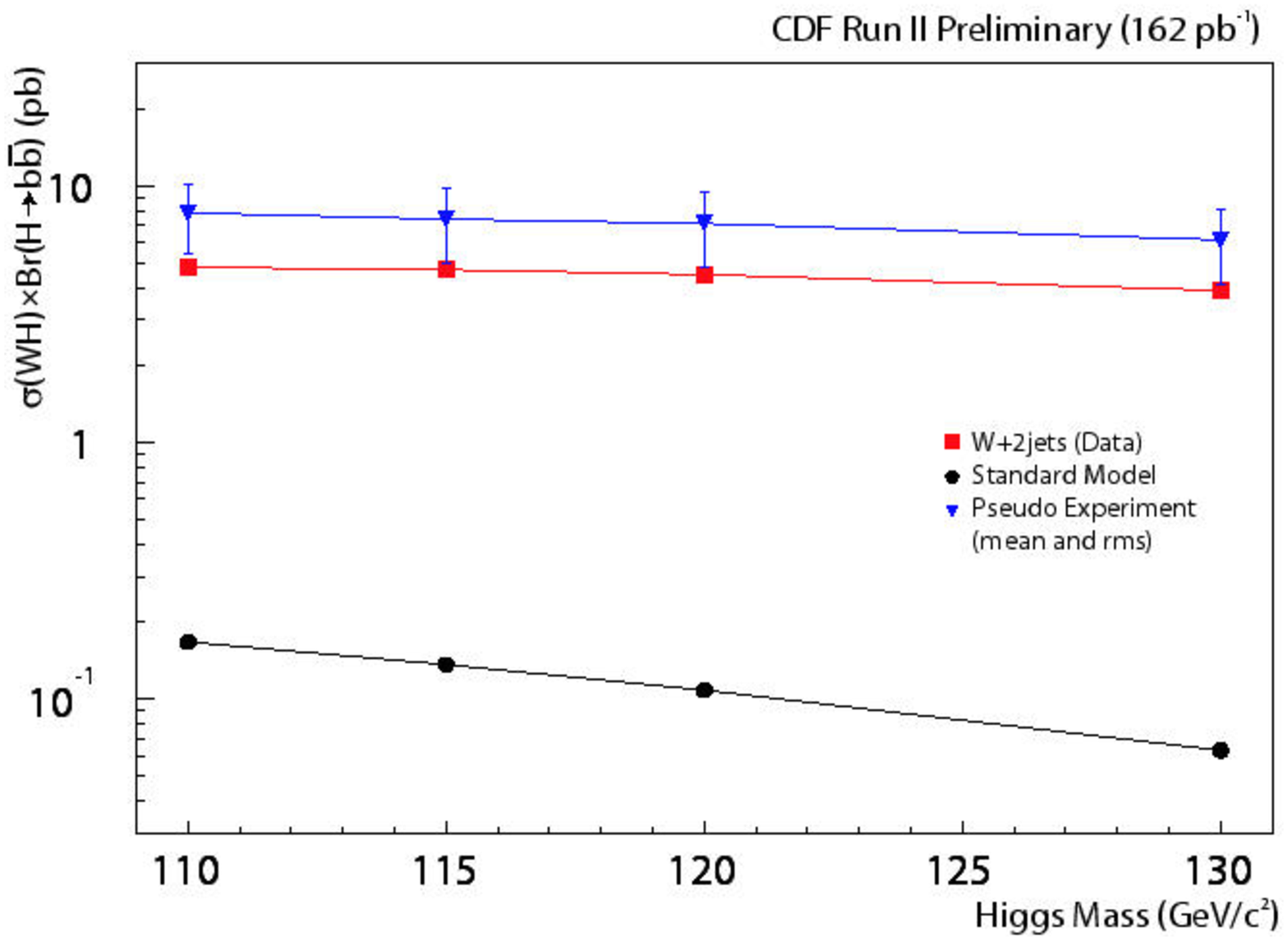,height=60mm}
\hskip 5mm
\epsfig{figure=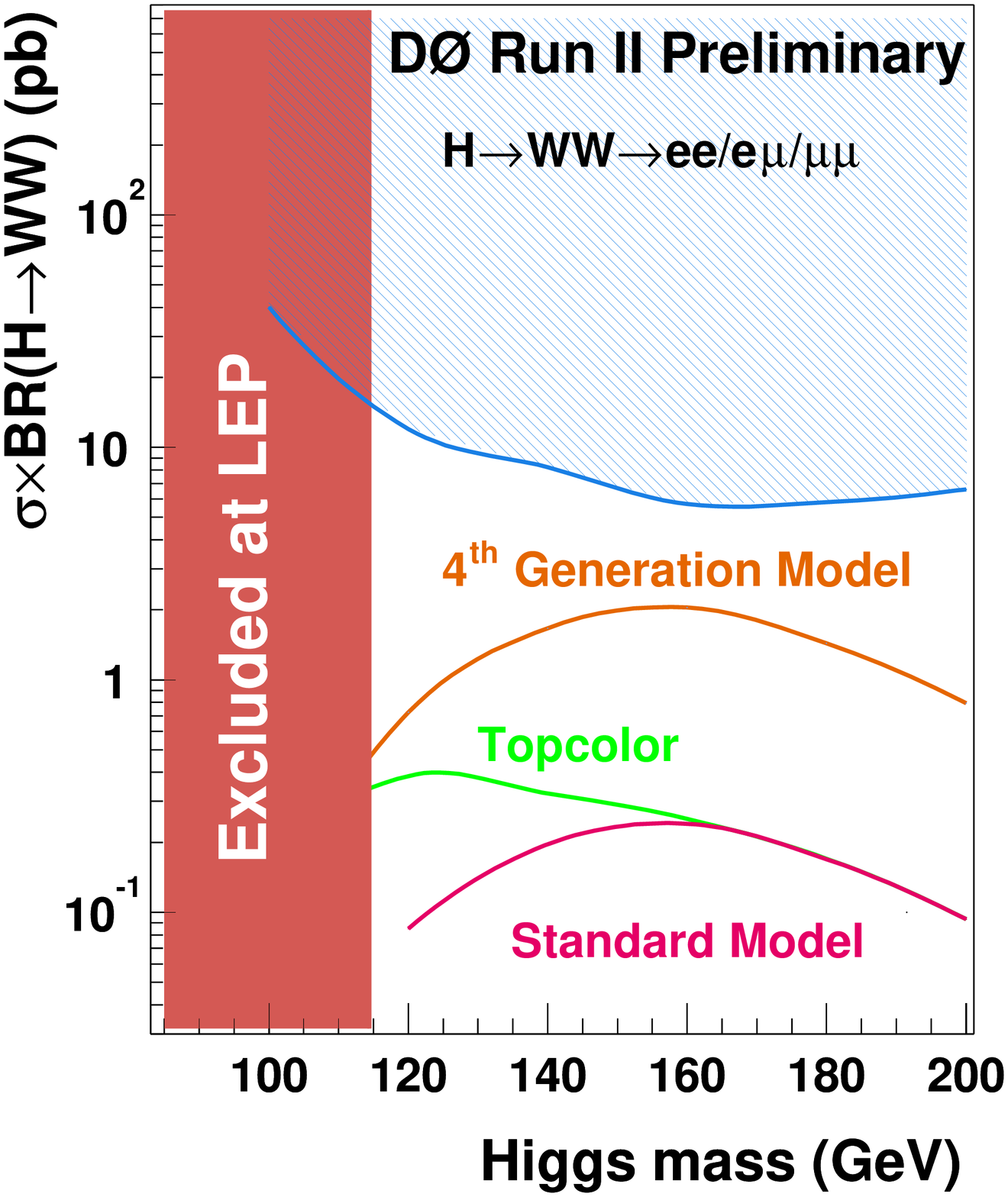,height=60mm}
}
\vskip -55mm
{\hskip 22mm} {\bf a)} {\hskip 80mm} {\bf b)}
\vskip 58mm
\caption{a) Limit on the $WH$ associated production cross section
times branching fraction,
$\sigma(WH)\times BR(W\rightarrow e\nu/\mu\nu)\times BR(H\rightarrow b\bar{b})$,
obtained by CDF; b) The SM Higgs to $WW$ production cross section
excluded by D\O\ .
\label{fig:cdfd0_smhiggs}}
\end{figure}

Searches for Non-SM Higgs at Tevatron were also reviewed~\cite{hughes}. Although no evidence has been found so far, the results have already improved on those of Run I.

\section{Searches for New Phenomena}
\label{sec:np}

Searches for Supersymmetry (SUSY) at the Tevatron were reviewed~\cite{demina}. This popular extension of the Standard Model originally suggested over 25 years ago postulates the symmetry between fermionic and bosonic degrees of freedom. As a result a variety of hypothetical particles is introduced. Presently available experimental data showed that if supersymmetric particles exist, they must be heavier than their Standard Model partners. In other words Supersymmetry is broken and theory suggests several possible scenarios mediated by gravitational or gauge interactions.  One possible exception is supersymmetric top quark (stop), which still has a chance to be lighter or of comparable mass as the top quark. With 2-4 $fb^{-1}$ of data Tevatron experiments will be able to extend the limit on stop mass above that of top quark or discover it and thus establish Supersymmetry. 

Meanwhile CDF and D\O\ collaborations analyzed up to 200 $pb^{-1}$ of the delivered data in search for different supersymmetry signatures, so far with negative results. Weak production of chargino and neutralino (superpartners of W and Z-bosons) can result in a distinct signature – three leptons accompanied by energy misbalance from escaping lightest suppersymmetric particles. By using a clever technique of lepton identification as an isolated track D\O\ was able to extend its sensitivity with respect to the previous analysis, which searched for three fully identified leptons. The excluded rate is still above that predicted by the theory and thus no mass range could be excluded. In contrast to weakly produced charginos and neutralinos, strongly produced superpartners of quarks and gluons (squarks and gluinos) have large production cross section. The signature of this process is a combination of energetic jets and missing energy. Though this signature suffers from substantial instrumental backgrounds D\O\ experiment extended the sensitivity region beyond that of Run I. CDF combined jets and missing energy signature with heavy flavor tagging to search for gluino cascades to bottom and sbottom quarks. With just 38 $pb^{-1}$ of data the sensitivity was extended beyond Run I because of the increased efficiency and acceptance of the silicon system. Gauge mediated SUSY breaking scenario can result in signatures enriched in photons. D\O\ explored this possibility and limited the scale of gauge mediated SUSY breaking to above 78.8 TeV. Finally, CDF used its newly acquired time-of-flight system to search for long lived heavy particles. 7 events were observed in 53 $pb^{-1}$ of data with expected number of events of 2.9$\pm$0.7(stat)$\pm$3.1(syst). The result was interpreted as a limit on stop cross section.   With more data to come these results will be updated and new channels will be added to fully exploit our chances for discovery.

Searches for non-SUSY extensions of the Standard Model at Tevatron were also reviewed~\cite{vilar}. Although no evidence for new physics has been found so
far, the results have already improved on those of Run I. The discovery
potential rises as the integrated luminosity increases, providing the
best opportunity for finding any evidence before the LHC starts. 

One of the main purposes of the LHC collider is to search for the
physics beyond the Standard Model. The discovery of superpartners of
ordinary particles, as expected in SUSY, would be a proof of the existence of new
physics. In the past years several studies have been carried on 
to understand the discovery reach of SUSY particles through inclusive 
studies. It has been demonstrated that, if supersymmetry 
exists at the electroweak scale, it could
hardly escape detection at LHC. Already 
with 1 fb$^{-1}$ of integrated luminosity, LHC should be able 
to discover squarks and gluinos if their masses do not exceed about 
1.3~TeV. With 100 fb$^{-1}$ the reach can be extended up to masses $m_{\squark}\sim
 m_{\gluino}\sim 2.5$ TeV. The inclusive 
studies show that the entire plausible domain of EW-SUSY parameter space 
for most probable value of tan$\beta$ can be probed.
The main problems is not to discover 
SUSY particles, but rather to perform a SUSY ``spectroscopy'' to 
determine masses and properties and, eventually, to probe 
different models. 
The ATLAS and CMS~\cite{tricomi} collaborations reported on new studies made in order to understand the detector capabilities to perform exclusive reconstructions.

\section{Conclusions}
\label{sec:conclusions}

Many interesting new results were presented at the Moriond QCD conference this year. These exciting results in variety of areas show how much progress was made over the past year in understanding the basic forces and building blocks of nature. With accelerators worldwide performing well and expected to deliver at an even higher level in the next several years, High Energy Physics is on the right path to a bright future.

\section{Acknowledgements}
\label{sec:acknowledgements}

I like to thank the conference organizers for a very productive and pleasant week. In particular, I wish to thank Tran Thanh Van and Etienne Auge for the invitation to present the summary talk and for all the help in assembling the material for this presentation. I also want to thank many of my colleagues at Moriond QCD for valuable discussions and assistance in putting together this contribution. In addition, I thank Tom Ferbel, Paul Grannis, and John Womersley for careful reading of this manuscript.

\section*{References}

\begin{thebibliography}{99}

\bibitem{fixed}
LEPS \coll, T.~Nakano \etal Phys.~Rev.~Lett. {\bf 91} (2003)
012002; \\ SAPHIR \coll, J.~Barth \etal Phys.~Lett. {\bf B~572}
(2003) 127;  \\ CLAS \coll, V.~Kubarovsky \etal Phys.~Rev.~Lett.
{\bf 91} (2003)~252001; CLAS \coll, V.~Kubarovsky \etal 
Phys.~Rev.~Lett. {\bf 92} (2004) 032001. Erratum; ibid, 049902. 

\bibitem{ks}
DIANA \coll, V.V.~Barmin \etal Phys.~Atom.~Nucl. {\bf 66} (2003)
1715; \\ A.E.~Asratyan, A.G.~Dolgolenko, M.A.~Kubantsev, Preprint
\mbox{hep-ex/0309042}, 2003; \\ SVD \coll, A.~Aleev \etal Preprint \mbox{hep-ex/0401024}, 2004; \\ HERMES \coll, A.~Airapetian \etal Phys.~Lett. {\bf B~585} (2004) 213; \\ COSY-TOF \coll, M.~Abdel-Bary \etal Preprint
\mbox{hep-ex/0403011}, 2004.

\bibitem{NA49}
NA49 \coll, C.~Alt \etal Phys.~Rev.~Lett. {\bf 92} (2004) 042003.

\bibitem{geiser}A. Geiser, {\it for the ZEUS Collaboration}, these proceedings.
\bibitem{wengler} T. Wengler, {\it for the LEP Collaborations}, these Proceedings.
\bibitem{tsusa}T.\v Su\v sa, {\it for the NA49 Collaboration}, these proceedings.
\bibitem{lipka}K. Lipka, {\it for the H1 and ZEUS Collaborations}, these proceedings.
\bibitem{mikami}Y. Mikami, {\it for the Belle Collaboration}, these proceedings.
\bibitem{calderini}G. Calderini, {\it for the \BABAR Collaboration}, these proceedings.
\bibitem{campanelli} M. Campanelli, {\it for the CDF and D\O\ Collaborations}, these proceedings.
\bibitem{denterria} D.~d'Enterria, {\it for the RHIC Collaborations}, these Proceedings.
\bibitem{fwang} F. Wang, {\it for the RHIC Collaborations}, these Proceedings.
\bibitem{liu} M.X. Liu, {\it for the RHIC Collaborations}, these Proceedings.
\bibitem{botje} M. Botje, {\it for the NA49 Collaboration}, these Proceedings.
\bibitem{sitta} M. Sitta, {\it for the NA50 Collaboration}, these Proceedings.
\bibitem{virgili} T. Virgili, {\it for the NA57 Collaboration}, these Proceedings.
\bibitem{arnaldi} R. Arnaldi, {\it for the NA60 Collaboration}, these Proceedings.
\bibitem{dainese} A.~Dainese, {\it for the ALICE Collaboration}, these Proceedings.
\bibitem{schoeffel}L. Schoeffel, {\it for the H1 and ZEUS Collaborations}, these proceedings.
\bibitem{anulli} F. Anulli, {\it for the \BABAR Collaboration}, these Proceedings.
\bibitem{zei} R. Zei, {\it for the NOMAD Collaboration}, these Proceedings.
\bibitem{elschenbroich} U. Elschenbroich, {\it for the HERMES  Collaboration}, these Proceedings.
\bibitem{kucharczyk} M. Kucharczyk, {\it for the LEP Collaborations}, these Proceedings.
\bibitem{denig} A. Denig, {\it for the KLOE Collaboration}, these Proceedings.
\bibitem{zavertyaev} M. Zavertyaev, {\it for the HERA-B Collaboration}, these Proceedings.
\bibitem{padley} P. Padley, {\it for the CDF and D\O\ Collaborations}, these proceedings.
\bibitem{hesketh} G. Hesketh, {\it for the CDF and D\O\ Collaborations}, these proceedings.
\bibitem{mcnulty} R. McNulty, {\it for the CDF and D\O\ Collaborations}, these proceedings.
\bibitem{yuan} C.~Z.~Yuan, {\it for the BES Collaboration}, these Proceedings.
\bibitem{napolitano} J. Napolitano, {\it for the CLEO Collaboration}, these Proceedings.
\bibitem{vaandering} E. Vaandering, {\it for the FOCUS Collaboration}, these Proceedings.
\bibitem{nedden} M. zur Nedden, {\it for the HERA-B Collaboration}, these Proceedings.
\bibitem{melzer} I. Melzer-Pellman, {\it for the H1 and ZEUS Collaborations}, these proceedings.
\bibitem{shepherd} M. R. Shepherd, {\it for the CLEO Collaboration}, these Proceedings.
\bibitem{duboscq} J.E. Duboscq, {\it for the CLEO Collaboration}, these Proceedings.
\bibitem{smith} J. Smith, {\it for the \BABAR Collaboration}, these Proceedings.
\bibitem{zhang} J. Z. Zhang, {\it for the Belle Collaboration}, these proceedings.
\bibitem{yip} K. Yip, {\it for the CDF and D\O\ Collaborations}, these proceedings.
\bibitem{ia} I.~Iashvili, {\it for the CDF and D\O\ Collaborations}, these proceedings.
\bibitem{cerrito} L.~Cerrito, {\it for the CDF and D\O\ Collaborations}, these proceedings.
\bibitem{juste} A.~Juste, {\it for the CDF and D\O\ Collaborations}, these proceedings.
\bibitem{roth} S.~Roth, {\it for the LEP-EW Working Group}, these proceedings.
\bibitem{avto}A.~Kharchilava,  {\it for the CDF and D\O\ Collaborations}, these proceedings.
\bibitem{hughes} R. Hughes,  {\it for the CDF and D\O\ Collaborations}, these proceedings.
\bibitem{demina}R. Demina, {\it for the CDF and D\O\ Collaborations}, these proceedings.
\bibitem{vilar}R. Vilar, {\it for the CDF and D\O\ Collaborations}, these proceedings.
\bibitem{tricomi} A. Tricomi, {\it for the ATLAS and CMS Collaborations}, these proceedings.

\end{thebibliography}

\end{document}
